\documentclass[%
  prc,10pt,
superscriptaddress, 
showpacs,preprintnumbers,
nofootinbib,
amsmath,amssymb,
aps
]{revtex4-1}

\usepackage{graphicx}
\usepackage{dcolumn}
\usepackage{bm}
\usepackage{color}

\def\k{{\boldsymbol k}}
\def\K{{\boldsymbol K}}
\def\P{{\boldsymbol P}}
\def\CM{_{\rm CM}}

\def\simge{\mathrel{
   \rlap{\raise 0.511ex \hbox{$>$}}{\lower 0.511ex \hbox{$\sim$}}}}
\def\simle{\mathrel{
   \rlap{\raise 0.511ex \hbox{$<$}}{\lower 0.511ex \hbox{$\sim$}}}}
\def\bigs{\mathrel{
   \rlap{\raise 0.531ex \hbox{$>$}}{\lower 0.531ex \hbox{$<$}}}}

\begin{document}

\title{Radiative hadronization: Photon emission at hadronization from quark-gluon plasma}

\author{Hirotsugu Fujii}
\email{hfujii@phys.c.u-tokyo.ac.jp}
\affiliation{
Institute of Physics, University of Tokyo, Komaba, Tokyo 153-8902, JAPAN
}%
\author{Kazunori Itakura}
\email{itakura$\_$kazunori@nias.ac.jp}
\affiliation{
Nagasaki Institute of Applied Science, Nagasaki 851-0193, JAPAN
}
\affiliation{%
KEK Theory Center, IPNS, KEK, Tsukuba 305-0801, JAPAN} 

\author{Katsunori Miyachi}%
\email{miyachi@hken.phys.nagoya-u.ac.jp}
\affiliation{
  Department of Physics, Nagoya University, Nagoya 464-8602, JAPAN
}%

\author{Chiho Nonaka$^{4,}$}%
\email{nchiho@hiroshima-u.ac.jp}
\affiliation{
Physics Program, Graduate School of Advanced Science and Engineering, Hiroshima University, 
Higashi-Hiroshima 739-8526, JAPAN}
\affiliation{
Kobayashi-Maskawa Institute for the Origin of Particles and the Universe (KMI), Nagoya University, Nagoya 464-8602, JAPAN \vspace{3mm}
}

\date{\today}

\begin{abstract}
We investigate photon emission at the hadronization stage from a quark-gluon plasma created in relativistic heavy-ion collisions.
A recombination-model picture suggests that a quark and an antiquark bind into a meson state
in hadronization, which would apparently violate the energy conservation if there is nothing else involved.
We consider here a hadronization process where the recombination accompanies a photon emission.
This is an analog of the ``{\it radiative recombination}" known in plasma physics,
such as $e^- + p^+ \to {\rm H}^0 +\gamma$, which occurs when an electromagnetic plasma goes back to a neutral atomic gas.
The ``radiative hadronization" picture will bring about 
  (i) an enhancement of the photon yield, 
  (ii) significant flow of photons similar to that of hadrons, and
(iii) the photon transverse momentum ($p_T$) distribution with a thermal profile whose effective temperature is given by blue-shifted temperature of quarks.
Here as a simplest and phenomenological realization of the radiative hadronization, we modify the recombination model to involve a photon emission 
and evaluate the photon yield with this modified model.
Adding this contribution to the direct photon yield along with thermal photon contribution calculated with a hydrodynamic model
and a parametrized contribution of prompt photons,
we study the $p_T$ spectrum and elliptic flow of the photons produced in heavy-ion collisions at RHIC and LHC energies.
\end{abstract}


\maketitle


\section{Introduction}

Direct photon production is of special importance in relativistic heavy-ion collisions (for reviews, see \cite{Gale:2009gc,Chatterjee:2009rs}). Since photons couple with the quark-gluon plasma (QGP) only through the electromagnetic interaction, their mean free path is much longer than the dimension of the reaction zone and they can escape from it carrying information on the QGP at the instance of their emission. These photons are called thermal photons, whose observation is regarded as indirect evidence of the formation of hot thermalized states of quarks and gluons in relativistic heavy-ion collisions \cite{Shuryak:1978ij,Kapusta:1991qp}. There are still other various photon sources over the time evolution in a heavy-ion collision event in addition to thermal photons from the QGP phase. For example, at the very initial stage, partonic hard collisions (such as the gluon Compton scattering $gq\to \gamma  q$ and quark annihilation $q\bar q \to \gamma g$) produce photons, which are called prompt photons \cite{Owens:1986mp,Turbide:2007mi,Klasen:2013mga}. At the late stage after hadronization, scattering processes between hadrons also produce photons \cite{Kapusta:1991qp,Holt:2015cda}. Thus, identifying and quantifying photon sources become a vital issue for extracting direct information on QGP from the photon observation.

Measurements of the photons in relativistic heavy-ion collisions have been performed at both RHIC and LHC, and the results of transverse momentum ($p_T$) distributions $dN^\gamma/dp_\perp^2 dy$ and elliptic flow coefficients ($v_2^\gamma$) at mid-rapidity $y\sim 0$ have been reported \cite{Adare:2014fwh,Adare:2015lcd,Adam:2015lda,Lohner:2012ct} (see also \cite{David:2019wpt} for recent experimental review). From the exponential slopes of the $p_T$ distributions, ``{\it effective temperatures}'' of produced photons at RHIC ($\sqrt{s_{NN}}=200~$GeV) and LHC ($\sqrt{s_{NN}}=2.76~$TeV) are found to be $T_{\rm eff}\sim 230~$MeV and $T_{\rm eff}\sim 300$~MeV, respectively. Very interestingly, the elliptic flow $v_2^\gamma$ of the direct photon distribution is found be as large as that of hadrons. Notice that although the ``decay photons'' produced in hadron decays (mainly due to $\pi^0\to 2 \gamma$) have been subtracted from the total yield to obtain the direct photon yield, it is still a mixture of photons from various sources. Despite many theoretical attempts to reproduce these experimental results, any theoretical model so far seems to be incapable of explaining the photon data adequately. For example, hydrodynamic models which explain the hadronic spectra and anisotropic flow very well, tend to underpredict the amplitude of photon elliptic flow (e.g., see \cite{Chatterjee:2021gwa}). In particular, even an up-to-date hydrodynamic model calculation with various possible effects included still underestimates the photon yield \cite{Paquet:2017wji,Gale:2021emg}. Other model calculations such as the blast-wave type fireball model and the ideal hydrodynamic model also give rise to smaller thermal photon yield and elliptic flow than the experimental data by PHENIX an ALICE \cite{vanHees:2014ida}. Furthermore, the parton-hadron-string dynamics (PHSD) model shares the same tendency in direct photon yield and elliptic flow \cite{Linnyk:2015tha}. Meanwhile, the pre-equilibrium Glasma stage is discussed in Refs.~\cite{Berges:2017eom,Monnai:2019vup} as a possible photon production source, but its contribution is also likely to suppress the elliptic flow.

This situation is referred to as the “direct photon puzzle”.
The difficulty comes from two seemingly contradictory aspects of the photon data, large yields and strong collective flow. The large yield could be attributed to the early stage of the evolution with higher temperatures, while the strong collective flow prefers large photon emission at the later stage when momentum anisotropy of QGP is well developed. Therefore, it is not easy to explain these two points within a single theoretical model. However we should note that there is a reservation about the experimental results because the large yield of photons measured by PHENIX has not been confirmed by STAR.

Given this situation, it should be very worthwhile to examine another source of photons which has been overlooked so far and is inherent to late stages of the QGP time evolution. This brought us to think of a possibility of photon radiations at hadronization of QGP, which is, in fact, natural from the viewpoint of ordinary electromagnetic plasmas. It is well-known that an electromagnetic plasma radiates numerous photons when it goes back to an atomic gas. This process is called the ``radiative recombination" and is seen in various astrophysical situations (for an overview see \cite{graham2012recombination, hahn1997electron}. This is natural because photon emission is advantageous for satisfying energy conservation in the formation of bound states (electrically neutral atoms). We can expect similar phenomena occur in the hadronization processes, where valence quarks and antiquarks form bound states. In the present paper, we formulate photon production at the hadronization stage in analogy with the radiative recombination in electromagnetic plasmas and investigate properties of the produced photons. Photon production during the QCD phase transition and hadronization has been attracting much attention indeed, and has been discussed in different frameworks, for example, in \cite{vanHees:2014ida,Campbell:2015jga,Young:2015adw}.

The present paper is organized as follows. In the next section, we explain the radiative recombination in ordinary electromagnetic plasma and comment on similarities to and differences from the QGP. Then in Sec.~III, we provide the basic theoretical framework for radiative recombination applied to hadronization.
Numerical results are presented in Sec.~IV, where we include the thermal photon contributions obtained by a hydrodynamic simulation to compare the results to the observed data.
Discussions and summary are given in Sec.~V.  

In Appendix A, we present photon production formulas in our radiative recombination model.
In Appendix B, we give a brief description of thermal photon calculation in a hydrodynamic model \cite{Miyachi-Nonaka}.
The model parameter dependence of our results is discussed in Appendix C.

\section{Radiative recombination in electromagnetic plasmas}

In this section, we explain what is known about radiative recombination in ordinary plasmas and in other relevant processes, which are helpful to understand how the radiative recombination in QGP could be formulated. The primary and simplest example is that of an electromagnetic plasma made of electrons and protons. When the temperature is decreased, the plasma decays and neutralizes into a hydrogen gas through the microscopic mechanism of the radiative recombination $e^-+p^+\to {\rm H}^0 +\gamma$. Emission of a photon compensates for the energy difference between the initial continuum state and the final bound state. The ordinary plasma flashes when it decays. This process is well-known in plasma physics and is also called ``free-bound transition" \cite{fujimoto2004plasma}. The secondary example is the glow discharge which is seen as a typical picture of a plasma. The third example is radiative recombination in the early universe: There was a ``recombination era" and without treating the radiative recombination we would not be able to accurately evaluate the ``recombination temperature'' which is essential for understanding the cosmic microwave background (CMB) data \cite{weinberg2008cosmology}. The fourth example is gas nebulae, such as the Orion Nebula, which have beautiful radiations containing continuum spectra due to the radiative recombination \cite{osterbrock2006astrophysics}. Lastly, in addition to these examples induced by the electromagnetic interaction, we also have examples in nuclear reactions. We know that in the sun there are two important processes called the ``pp chain" and the ``CNO cycle," both of which include formation of bound states accompanied by a photon emission (such as $D + p \to {}^{3}{\rm He}+\gamma$ in the pp chain and $p+{}^{12}{\rm C}\to {}^{13}{\rm N}+\gamma$ in the CNO cycle) \cite{clayton1968principles}. All these examples indicate that radiation is naturally inherent to the formation of bound states, which suggests the possibility of photon emissions at hadronization, provided that hadronization can be modeled as a coalescence process of valence partons.

There are two key equations for the description of radiative recombination in electromagnetic plasmas \cite{rybicki2008radiative,weinberg2008cosmology}. One is the Kramers-Milne relation which relates recombination cross section to that of its inverse process (photo-ionization $\gamma + A \to e^- + A^+$) and corresponds to the detailed balance relation in thermal equilibrium states. The other is the Saha equation which gives the ionization ratio $X=n_{\rm ion}/(n_{\rm ion}+n_{\rm atom})$ between the atom and ion numbers, $n_{\rm atom}, n_{\rm ion}$, as a function of a temperature when the electrons, ions, and atoms are in thermal equilibrium. The Kramers-Milne relation is useful because the photo-ionization rate is easily measured by experiments, and the Saha equation applies to an ordinary plasma because one can control the lifetime of a plasma much longer than the microscopic time scale of radiation reaction. These relations are, unfortunately, not suitable to the hadronization from QGP in heavy-ion collisions since it should be treated as a nonequilibrium process. Hadronization occurs in a small lump of QGP at the time scale of the strong interaction, and therefore the photons are just emitted without re-absorption. To this situation we cannot use the relations which assume a system in bulk equilibrium.

When the density of an electromagnetic plasma is relatively high, another type of recombination will be possible. It is the ``three-body recombination" $2 e^- + A^+\to e^- + A$, where the energy conservation is satisfied by the spectator electron. 
If one defines the effective photon emission rate at some density, it will be reduced by the presence of the three-body, or in general, multi-body recombination. We may expect similar phenomena in the hadronization process. Formation of bound states without photon emission will be possible if the valence partons interact with other particles in the medium. We will be able to absorb this kind of effects into an effective recombination rate including density dependence. In fact, as we will discuss later,  since we treat the recombination rate as a parameter, we expect that such kind of effects are included in the overall normalization parameter which is determined to fit the experimental data. Alternatively, effects of multi-body recombination without a photon emission will be described by ``off-shell'' valence partons. However, within our framework in the present analysis, it is not easy to define off-shell partons in QGP. 

Lastly, we comment on a very similar process in $e^+e^-$ collisions.  Recall that the $e^+e^-$ collisions have been used to discover new particles by changing the invariant mass. Pronounced resonances such as $\rho, \omega, \phi$ and $J/\psi$ mesons are measured with energies {\it below} the invariant mass of the $e^+e^-$ system. This is called the {\it radiative return}, and is quite important for the analysis of the $R$-ratio around threshold and $(g-2)$ of leptons (for reviews, see \cite{Actis:2010gg, Druzhinin:2011qd}). Notice that the formation of a resonance below the $e^+ e^-$ invariant mass is accompanied by photon emissions and the typical radiative return is represented as 
$
e^+ + e^- \to \text{hadrons} + n\gamma\, ,
$
where the number $n$ of emitted photons is not necessarily one, $n\ge 1$. In particular, a clean process with a single meson and a single photon 
$
e^++e^-\to \text{meson} + \gamma
$ 
is experimentally measured. For example, $J/\psi$ production was observed at BaBar \cite{Aubert:2003sv}, and more recently $\chi_c$ and $\eta_c$ at BESIII \cite{Ablikim:2014hwn, Ablikim:2017ove}. The counterpart in a purely QED case such as $e^+e^-\to \mu^+\mu^-\gamma$ can be perturbatively calculated (though quite tedious), and one can study interplay between the initial state radiation and the final state radiation. On the other hand, hadron production suffers from ambiguity related to the coupling between the virtual photon and a composite hadron. For example, $e^++e^- \to \gamma^* \to \text{meson}+\gamma$ includes the transition of a virtual photon into a meson which may be phenomenologically described by the vector meson dominance.  Heavy quarkonium production will have less ambiguity, but the NRQCD formalism developed for the calculation of heavy-quarkonium production involves nonperturbative matrix elements (see for example, \cite{Chung:2008km, Sang:2009jc}).

There are Monte-Carlo generators for the radiative return called PHOKHARA  (for real photon emission) and EKHARA (for virtual photon emission) \cite{Czyz:2017veo}. These generators treat radiation by the vertices like $meson^*\to meson + \gamma\, ({\rm or}\ \gamma^*)$ which are given by effective lagrangian \cite{Czyz:2012nq}. Here, $meson^*$ could be a virtual (or off-shell) meson. This kind of picture will be useful in our problem. Another lesson from the radiative return is that the final state could involve several hadrons, typically light mesons such as pions and kaons. For example, final states with four mesons like $\pi^+ \pi^- K^+ K^-$ were extensively studied \cite{Actis:2010gg}. We expect similar multiple hadron production in the radiative hadronization. As we will discuss in the next section, we will adopt the ``Recombination model" and modify it so that it allows photon emission. This recombination model provides the number of produced mesons by the overlapping between a $q\bar q$ state and a meson state, and implicitly assumes a single meson production from a $q\bar q$ state. However, the multiple hadron production seen in the $e^+e^-$ collisions suggests that the one-to-one correspondence between a $q\bar q$ state and a meson should not exactly hold. Still, we may be able to effectively absorb such effects into the recombination rate. We should be aware that the overall recombination rate could contain many different physical effects. Having said all these suggestions and caveats, we are now ready for the problem of how to formulate the radiative recombination at hadronization.

\section{Radiative hadronization: formulation}

Hadronization is a nonperturbative process because it takes place around the critical temperature $T\sim T_c$ and the typical strong coupling $\alpha_s(T\sim T_c)$ is not small. It is also a nonequilibrium phenomenon in the sense mentioned in the previous section. One of the possible frameworks to describe the hadronization will be to work in an effective theory that includes both hadronic and constituent-quark degrees of freedom (see \cite{Young:2015adw} for an analysis in the quark-meson coupling theory). Within this framework, one can compute the hadron production cross sections similarly to the radiation return. We will, however, take an alternative approach. In fact, as already commented before, we know a simple and phenomenologically successful model for hadronization which is based on the coalescence of constituent quark degrees of freedom. It is the recombination  model  and we utilize it to describe the radiative hadronization. Below we first explain briefly the basic strategy of the recombination model, and then discuss how to modify it to include the photon emission.

\subsection{Recombination model}

The recombination/coalescence models provide a phenomenological description of hadronization for hadron production in the intermediate $p_T$ region ($2 \simle p_T \simle 5$~GeV/c) and give a natural explanation for intriguing phenomena such as the anomalous baryon/meson ratio and constituent quark number (CQN) scaling of the elliptic flow \cite{Fries:2003vb,Greco:2003xt,Molnar:2003ff,Hwa:2003bn} (for a review, see \cite{Fries:2008hs}). There are several recombination/coalescence models with some differences \cite{Fries:2003kq, Greco:2003mm,Hwa:2004ng}, and we will adopt the ReCo model that was developed by Duke group \cite{Fries:2003kq}.

The ReCo model starts by defining the number of hadrons that one can find in the quark/antiquark distributions. For example, the total number of mesons is defined as an overlap between the meson state $|M;\bm{P}\rangle$ and the reduced two-body density matrix $\hat \rho_{ab}$ which represents the partonic system with partons, $a$ and $b$, undergoing hadronization:
\begin{eqnarray}
N_M=\sum_{ab}\int \frac{d^3\bm{P}}{(2\pi)^3}\, \langle M; \bm{P}|\, 
\hat \rho_{ab} \, |M;\bm{P}\rangle    \, ,
\end{eqnarray}
where summation is taken over all the possible combinations of partons $a, b$ that have nonzero overlap with a mesonic state.
In this sense, the ReCo model simply projects the partonic picture of the QGP onto the hadron picture, and does not describe dynamical processes of hadron formation. However, since the formula includes  matrix elements like $\langle M;\bm{P}|\bm{r}_1,\bm{r}_2\rangle$ with $|\bm{r}_1,\bm{r}_2\rangle$ being a state having a quark at $\bm{r}_1$ and an antiquark at $\bm{r}_2$, it allows for an intuitive understanding of coalescence processes\footnote{Note however that this matrix element will be interpreted as a quark-antiquark component of a meson wavefunction.} like  $q\bar q\to M(\text{meson})$ and $qqq\to B(\text{baryon})$.

After some manipulations (see \cite{Fries:2003kq} for details), one obtains the momentum distribution of mesons made of partons $a$ and $b$ as 
\begin{equation}
E\frac{dN_M}{d^3\bm{P}}=C_M \int_\Sigma \frac{P\cdot u(R)}{(2\pi)^3}\int_0^1 dx\,  w_a(R;x\bm{P})\, \left|\phi^{}_M(x)\right|^2\, w_b(R;(1-x)\bm{P}) \, , 
\label{dN/dP}
\end{equation}
where $P^\mu=(E,\bm{P})$ is the four-momentum of the meson $M$, and $R$ is a four-vector specifying a point on the hypersurface $\Sigma$ where the hadronization takes place, and $u^\mu (R)$ is a unit vector orthogonal to the hypersurface $\Sigma$ at $R$. The function $\phi_M(x)$ is the light-cone wavefunction of a meson with $x$ being the momentum fraction of one of the two quarks, and $w_a(R;x\bm{P})$ is a one-particle phase space distribution of parton $a$,

We assume the longitudinal boost-invariant expansion (Bjorken expansion) of the QGP so that the hadronization hypersurface $\Sigma$ has a constant longitudinal proper time $\tau=\sqrt{t^2-z^2}=\, $const and a point $R^\mu$ on it is specified as
$$
R^\mu = (t,x,y,z)=(\tau \cosh \eta,\, \rho \cos \phi,\, \rho \sin \phi ,\,  \tau \sinh \eta)\, 
$$
with the space-time rapidity $\eta$, the transverse radial coordinate $\rho$, and the azimuthal angle $\phi$. Then the forward normal vector $u^\mu(R)$ orthogonal to the hypersurface $u\cdot dR|_{\tau= \rm const.}  =0$, is given by $u^\mu(R)=(\cosh \eta,0,0,\sinh \eta)$.

Regarding $\phi_M(x)$, we expect that our results are insensitive to its details and take $|\phi_M(x)|^2=\delta(x-1/2)$ for analytic evaluation and $\phi_M(x) =\sqrt{30}\, x(1-x)$ for numerical evaluation. Notice that both examples of the wavefunction have a peak at $x=1/2$, and therefore $w_a(R,\tfrac{1}{2}\bm{P})w_b(R,\tfrac{1}{2}\bm{P})\sim {\rm e}^{-P/T}$ gives rise to the dominant configuration in the $x$ integration. 
The overall factor $C_M$ counts state degeneracy. The corresponding formula for baryon production has the factor $C_B$ and the wavefunctions of the three-quark state should have a peak around $x=1/3$, and therefore $w_a(R,\tfrac{1}{3}\bm{P})w_b(R,\tfrac{1}{3}\bm{P})w_c(R,\tfrac{1}{3}\bm{P})\sim {\rm e}^{-P/T}$. Thus, this model predicts that the ratio of the proton to the pion yield at the common $P_T$ at mid-rapidity, $R_{p/\pi}\equiv \frac{dN_p}{d^2P_Tdy}\big{/}\frac{dN_{\pi^0}}{d^2P_Tdy}$, is essentially given by a ratio $C_B/C_M$, which amounts to $\sim 2$. This is indeed consistent with the experimental result known as the anomalous baryon/meson ratio, which is in contrast to the expectation from parton fragmentation processes in perturbative QCD, $R_{p/\pi}\sim 0.2$.

For $w_a(R;x\bm{P})$, 
we assume that at the onset of hadronization the quarks/antiquarks are in local thermal equilibrium with a fluid flow velocity $\bar v^\mu(R)$, which we parametrize as 
\begin{equation}
\bar v^\mu(R)=(\cosh \eta^{}_L \cosh \eta^{}_T,\, \sinh \eta^{}_T \cos \phi, \, \sinh \eta^{}_T \sin \phi,\, \sinh \eta^{}_L \cosh \eta^{}_T)\, , \label{normalized_flow_vector}
\end{equation}
where $\eta^{}_L$ and $\eta^{}_T$ are longitudinal and transverse flow rapidities, respectively. This four-velocity is normalized as $\bar v_\mu \bar v^\mu=1$.
In the Bjorken expansion, which we assume in this work, the longitudinal flow rapidity $\eta_L$ is identified with the space-time rapidity $\eta$, i.e., $\eta^{}_L=\eta$, and the longitudinal flow velocity is $v_L=\tanh \eta = z/t$. On the other hand, the transverse rapidity $\eta_T$ is related to the transverse flow velocity $v^{}_T=\sqrt{v_x^2+v_y^2}$ by
\begin{equation}
  v^{}_T=\tanh \eta^{}_T \, ,
  \label{transverse_velocity}
\end{equation}
at mid-rapidity ($\cosh \eta_L = 1$).
We assume that $v_T$ is independent of the transverse radius $\rho$ for computational simplicity.

By using this flow velocity $\bar v^\mu(R)$, we take the one-body phase space distribution of parton $a$ as 
\begin{equation}
w_a(R;p)=\gamma_a \, {\rm e}^{-p\cdot \bar v(R)/T} {\rm e}^{-\eta^2 /2\Delta^2}
f(\rho, \phi)\, ,
\label{phase_space_dist}
\end{equation}
where $\gamma_a$ is the fugacity factor of parton $a$.
The factor ${\rm e}^{-\eta^2 /2\Delta^2} f(\rho, \phi)$ describes the spatial profile of the hot medium.
For central collisions, one may simply assume a constant profile $f(\rho,\phi)=\theta(\rho_0-\rho)$
within the transverse radius $\rho_0$ of the fireball at the recombination time $\tau$.
In more general cases, the transverse profile will be adjusted to reproduce the collision-centrality dependence of observed meson yields.
Meanwhile, the $\eta$-dependence may be ignored as far as the mid rapidity region is concerned. These approximations will be used in analytic evaluations of the ReCo model below.

We have assumed that the quark momentum distribution in $w_a$ has the thermal profile
${\rm e}^{-p \cdot \bar v(R)/T}$
of temperature $T$, boosted by the collective flow $\bar v(R)$.
We introduce elliptic anisotropy in this momentum distribution by a weak modulation of the transverse flow rapidity $\eta_T$
around the mean $\overline{\eta}_T$:
\begin{align}
   \eta_T(\phi;p_T)=\overline{\eta}_T (1 - h(p_T)\cos 2\phi) \, .
\label{etaT_phi}
\end{align}
The modulation amplitude $h(p_T)$ is assumed to be $p_T$-dependent:
\begin{align}
    h(p_T)=\frac{\alpha}{1+(p_T/p_0)^a}
\label{hpt-profile}
\end{align}
with $\alpha=(1-r)/(1+r)$ fixed by the transverse aspect ratio $r$ of the almond-shape collision zone,
and with $a$ and $p_0$ the constant parameters controlling the momentum dependence.
We then calculate the parton distribution (\ref{phase_space_dist}),
to find the elliptic flow coefficient $v_2^a(p^{}_T)$ of parton $a$ defined by 
\begin{eqnarray}
w_a(R;p)=\overline w_a(R;p)\Big(1+2v_2^a(p^{}_T)\cos 2\phi \Big)\, ,
\end{eqnarray}
where $\overline w_a(R;p)$ is the part independent of the azimuthal angle $\phi$.
Inserting this distribution into the formula (\ref{dN/dP}), we can compute the elliptic flow coefficient for
a meson transverse momentum ($P_T$) distribution at mid-rapidity:
\begin{equation}
  v_2^M(P_T)\equiv \langle \cos 2\Phi \rangle^{}_{P_L=0}
  =\frac{{\displaystyle \int d\Phi \cos 2\Phi \left(\frac{dN_M}{d^2P_TdP_L}\Big|_{P_L=0}\right)} }
        {{\displaystyle \int d\Phi  \left(\frac{dN_M}{d^2P_TdP_L}\Big|_{P_L=0}\right)}}\, ,
\end{equation}
where $\Phi$ is the azimuthal angle of the produced meson momentum.
If we adopt the $\delta$-function approximation for the light-cone wavefunction $|\phi_M(x)|^2=\delta(x-1/2)$
and assume a universal elliptic flow coefficient for all quark flavors,
$v_2^q(p_T)\equiv v^a_2(p_T)$,
then the elliptic flow of the meson momentum distribution is analytically evaluated as 
\begin{eqnarray}
v_2^M(P_T)=\frac{2\, v^q_2(\frac12 P_T)}{1+2\, v^q_2(\frac12 P_T)^2}
\, ,
\end{eqnarray}
which simplifies further for $v^q_2(p_T)\ll 1$ to
\begin{eqnarray}
v_2^M(P_T)\simeq 2v^q_2(P_T/2)\, .
\end{eqnarray}
Similarly, for baryons one finds 
\begin{eqnarray}
v_2^B(P_T)\simeq 3v^q_2(P_T/3)\, .
\end{eqnarray}
Therefore, the elliptic flow coefficient of a hadron with $n$ constituents satisfies
the following scaling:
\begin{eqnarray}
v_2^h(P_T)\simeq nv^q_2(P_T/n)\, .
\label{eq:CQNscaling}
\end{eqnarray}
If we plot $v_2^h(P_T)/n$ as a function of $P_T/n$ for various hadrons, the results will collapse into a single curve which is determined by the quark  elliptic flow coefficient $v_2^q$. This ``CQN scaling" is indeed observed in experimental data, and is regarded as one of the evidences for the quark recombination in hadronization and also for formation of a thermalized QGP in relativistic heavy-ion collision experiments. Nevertheless, as is obvious from the above derivation of the scaling, it appears only in an idealized situation and a certain deviation is expected from the scaling limit even when the recombination mechanism dominates in hadronization. Such deviations will have different sources, from which we can extract physical information on hadronization.

As we emphasized before, the recombination model describes the meson (baryon) formation as a 2-to-1 (3-to-1) process. If we take this literally, it implies that the model violates the energy conservation law because the invariant mass of the initial state with two (three) partons cannot be the same as the energy of a bound state. In Ref.~\cite{Fries:2003kq}, it is argued that the energy conservation would be preserved by the effects of interactions with the medium (which generates a width in parton dispersion) and that omission of explicit treatment of such effects would not significantly affect the bulk features of hadron production.
The medium effect on the recombination reminds us of the three-body recombination in electromagnetic plasmas, where one of the three is just a spectator to keep the energy conservation.

But, at the same time, we notice another process which satisfies the energy conservation law in electromagnetic plasmas. It is the radiative recombination. The counterpart process in hadronization should be also possible.\footnote{While it is possible to make energy be conserved within a dynamical model including the effects of resonances as developed in \cite{Ravagli:2007xx}, we will present a different picture. } In the next subsection we will discuss how to modify the ReCo model so that it describes the radiative hadronization.

\subsection{Radiative hadronization model}

We modify the ReCo model so as to allow for a photon emission, calling it the radiative ReCo model.
Then, the meson formation process becomes a 2-to-2 process,
\begin{equation}
  q+\bar q \to M+\gamma
  \, ,
  \label{radiative_hadronization}
\end{equation}
which fulfills the energy conservation law. A similar modification for baryon formation with a photon emission: $qqq\to B + \gamma$ should be also possible.

\begin{figure}[t] 
\begin{center}
\vspace*{-1mm}
   \includegraphics[width=0.7\hsize]{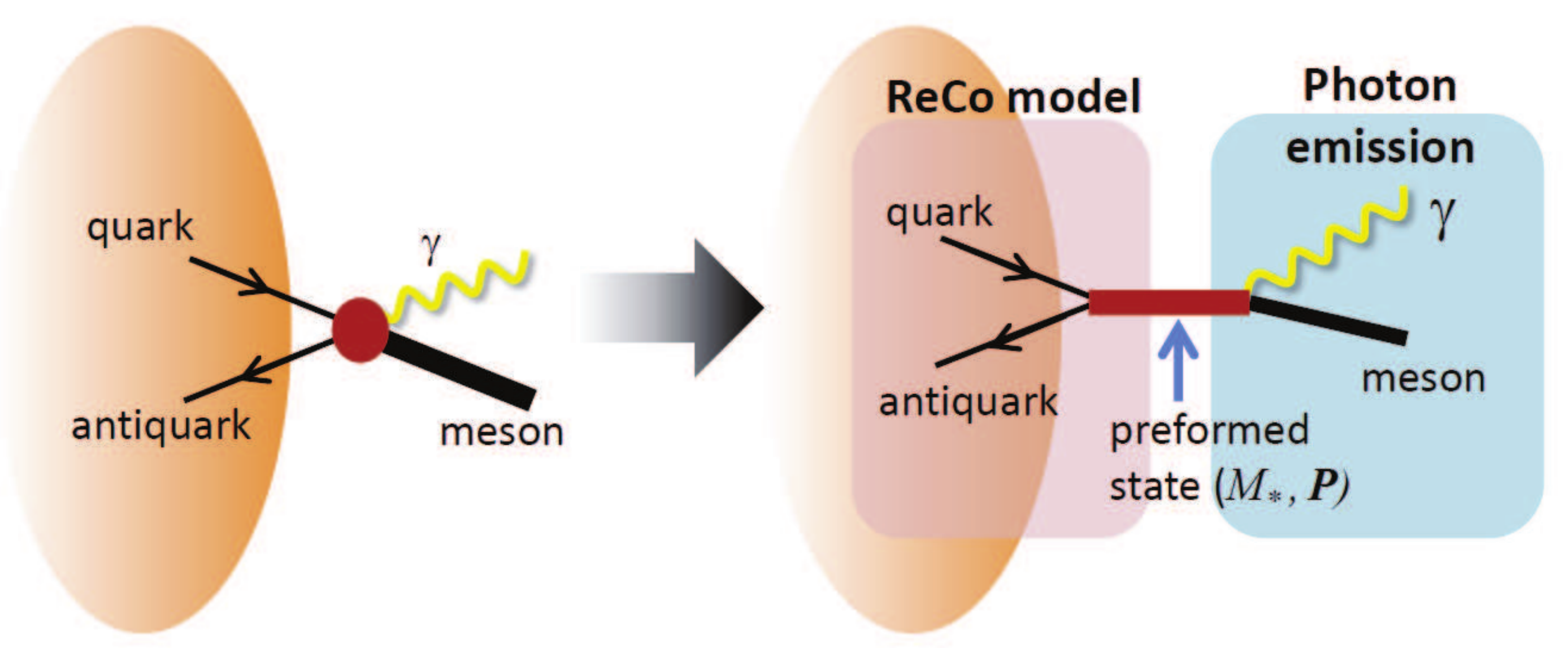}
\end{center}
\caption{Radiative ReCo model with a photon emission.}
\label{fig:ReCo}
\end{figure}

In our radiative ReCo model, we re-interpret the original ReCo model \cite{Fries:2003kq} as a tool for picking up a ``preformed state'' consisting of two partons and assume the preformed state emits a photon to form a bound state (see Fig.~\ref{fig:ReCo}). Notice that we do not consider this preformed state as any physical resonance but just as an intermediate state in radiative meson production.

The preformed state consisting of two partons with momenta $p_1^\mu=(E_1, \bm{p}_1)$ and $p_2^\mu=(E_2, \bm{p}_2)$,
has the invariant mass $M_*$ and momentum $\bm{P}$ determined by
\begin{eqnarray}
E \equiv \sqrt{M_*^2+\bm{P}_{}^2}&=&\sqrt{m_1^2+\bm{p}_1^2}
+\sqrt{m_2^2+\bm{p}_2^2}\, ,\\
\bm{P}&=&\bm{p}_1 + \bm{p}_2\, ,
\end{eqnarray}
where $m_1$ and $m_2$ are the constituent quark masses.
The invariant mass $M_*$ is a function of $\bm{p}_1$ and $\bm{p}_2$ and is evidently larger than $m_1+m_2$:
$$
M_*\ge m_1+m_2\, .
$$
On the other hand, in the constituent quark picture, the mass of the bound state $M$ should be smaller than $m_1+m_2$ by the binding energy: $M < m_1+m_2$. The surplus energy of $M_* - M>0$ should be carried away by a photon emission here.
Then, the number of the photons emitted in the formation of a meson $M$ reads
\begin{equation}
E_\gamma \frac{d N_\gamma}{d^3k}=\kappa \int dM_*\, \varrho(M_*)\int d^3P 
\left(\frac{dN_{M_*}}{d^3P}\right)\left(E_\gamma \frac{dn_\gamma(k; M_*,P)}{d^3k }\right) \, ,
\label{photon_distribution}
\end{equation}
which means that it is given by the product of the number of preformed states and the photon distribution emitted from a preformed state. We explain each ingredient below.

First of all, ${dN_{M_*}}/{d^3P}$ is the number of the preformed states which is characterized by momentum ${\bm P}$ and invariant mass $M_*$. We evaluate this part by the original ReCo model. Although we do not know the light-cone wavefunction of the preformed state, we expect that the results are insensitive to its details as we already commented concerning the original ReCo model. We use the same light-cone wavefunctions as in the original ReCo model. 

Next, $E_\gamma {dn_\gamma(k; M_*,P)}/{d^3k}$ corresponds to the photon distribution emitted from the preformed state moving with the momentum $\bm{P}$. We adopt a tree picture $M_*\to M+\gamma$ which is the leading order with respect to QED coupling $\alpha_{\rm em}$. We assume no specific polarization of the preformed states, and the photon distribution is treated as isotropic in the rest frame of the preformed state. More explicitly, we use the following photon distribution:
\begin{equation}
  E_\gamma \frac{dn_\gamma}{d^3 k}
  = \frac{1}{4\pi k_0}\, \delta (E_{\gamma \CM} - k_0)
  = \frac{M_*}{4\pi k_0}\, \delta (k \cdot P - k_0 M_*), 
\label{photon_CM_dist}
\end{equation}
where $E_{\gamma \CM}$ is the photon energy
and $k_0\equiv (M^2_*-M^2)/(2M_*)$ is the momentum magnitude of the photon and the meson in the center-of-mass (CM) frame of the preformed state ($M_*$-rest frame) with $M$ being the mass of the accompanying meson. The photon distribution is normalized as $\int dn_\gamma =1$.

We also introduced $\varrho(M_*)$ to represent an invariant mass  distribution of the preformed states of two partons. It should have a support for $M_*\ge m_1+m_2$. Considering the thermal distributions of quarks and antiquarks, we expect that the number of preformed states will rapidly decrease with increasing $M_*$. Therefore, we will replace it with the threshold contribution, {\it i.e.}, $\varrho(M_*)= \delta(M_*-(m_1+m_2))$ in this paper.

Finally and importantly, we comment on the overall factor $\kappa$ which is introduced to reflect other possible effects on radiative hadronization. Consider the recombination process mediated by a preformed state $q+\bar q \to M_* \to X$ where $X$ stands for any physical final states, including the radiative hadronization $X=M+\gamma$.
In general, however, $X$ can be multiple hadron states (with photons), as we discussed previously in relation to the radiation return in $e^+e^-$ collisions.
Once we compute all possible diagrams for the decay of $M_*$, we are able to define the ``branching ratio" for the radiative hadronization which corresponds to the factor $\kappa$. Since the transition probability for $M_*\to M+\gamma$ would be proportional to the QED coupling $\alpha_{\rm em}=1/137$, one may naively expect that $\kappa$ would be of the order of $\alpha_{\rm em}$. On the other hand, we also know empirically that the CQN scaling works very well up to several GeV of meson momentum $p_T$. This fact suggests that a single meson formation would be the dominant process over multiple meson production. If so, the single photon emission attached to this dominant process may have different value of $\kappa$ from the naive expectation.
Therefore, we leave the overall factor $\kappa$ as a parameter to be determined
by the experimental data.\footnote{One can consider gluon radiation,
  $M+g$, but the gluons are strongly interacting and will be re-absorbed into the medium.}

We remark here that the number of mesons is given by the same formula as eq.~\eqref{photon_distribution},
with the last factor replaced by the meson distribution emitted from the preformed state:
\begin{equation}
E_M \frac{d N_M^{\rm radReCo}}{d^3K}= \kappa \int dM_*\, \varrho(M_*)\int d^3P 
\left(\frac{dN_{M_*}}{d^3P}\right)
\left(E_M \frac{dn_M(K; M_*,P)}{d^3K}\right) \, ,
\label{meson_distribution}
\end{equation}
where $K^\mu$ is a momentum of the produced meson, satisfying $P_\mu=K_\mu + k_\mu^\gamma$. The distribution $dn_M/d^3K$ can be defined in a similar way to the photon case.

Recall that the original ReCo model naturally explains the CQN scaling.
In our modified ReCo model, on the other hand, the scaling should appear at the level of the distribution of preformed states $M_*$ in the integrand of eq.~(\ref{meson_distribution}) and
the meson production accompanying a photon emission may violate the CQN scaling to some extent.
We check this point both analytically and numerically in this paper.
But we emphasize here that the main contribution to hadron yield at around 2 GeV is given by the original ReCo model
and the radiative hadronization \eqref{meson_distribution} is a subdominant process of order $\kappa$ at most.
Moreover, since the photon carries away a fraction of the preformed-state momentum, the meson spectrum of the radiative hadronization is shifted to the lower momentum region and therefore at a given momentum $p_T$ its yield is more suppressed than the value simply expected by the factor $\kappa$.
However, we stress here that the radiative hadronization can give a significant contribution to photon production.

\subsection{Characteristics of the radiative hadronization}

In order to understand characteristics of the radiative hadronization, let us evaluate the momentum distributions of photons (\ref{photon_distribution}) and mesons (\ref{meson_distribution}) under certain approximations.

\subsubsection{Distribution of the preformed states}

We compute the number of preformed states by using the formula (\ref{dN/dP}) of the original ReCo model \cite{Fries:2003kq} and we recap the formulas of
ReCo model here. The transverse momentum spectrum of the preformed state at mid rapidity $\eta= 0$ is given by  \cite{Fries:2003kq}
\begin{equation}
E_{M_*}\left.\frac{dN_{M_*}}{d^3P}\right|_{\eta=0}\sim C_{M_*} M_{*T} \frac{\tau A_T}{(2\pi)^3} \; 2\gamma_a\gamma_b I_0\left(\frac{P_T\sinh \eta_T}{T_{\rm reco}}\right)
\int_0^1 dx|\phi_{M_*}(x)|^2 k_{M_*}(x,P_T)
\end{equation}
with 
\begin{equation}
k_{M_*}(x,P_T) \equiv K_1\left(\frac{\cosh \eta_T}{T_{\rm reco}}\left\{\sqrt{m_a^2+x^2P_T^2}+\sqrt{m_b^2+(1-x)^2P_T^2}\right\}\right)\, ,
\end{equation}
where $M_{*T}=\sqrt{P_T^2+M_*^2}$ is the transverse mass,
$A_T$ is the transverse area of the parton system at hadronization, representing the collision geometry, and $I_0(x)$ and $K_1(x)$ are the modified Bessel functions.
The parameter $T_{\rm reco}$ is the recombination temperature at which quark recombination $q+\bar q\to M_*$ occurs.
For analytic evaluation, we simply take $|\phi_{M_*}(x)|^2=\delta(x-1/2)$
and perform the integration over $x$ for pion production:
$$
\int_0^1 dx|\phi_{M_*}(x)|^2 k_{M_*}(x,P_T)\simeq K_1\left(\frac{\cosh \eta_T}{T_{\rm reco}}M_{*T}\right)\, ,
$$
where  $M_{*T}$ appears in the argument because we have taken $m_a=m_b=m$
and $M_*= 2m$ (recall that we adopt $\varrho(M_*)=\delta (M_*-(m_1+m_2))$). By using the asymptotic forms of the modified Bessel functions $I_0(z)\sim {\rm e}^z/\sqrt{2\pi z}$ and $K_1(z)\sim \sqrt{\pi/2z}\ {\rm e}^{-z}$ for large $z\gg 1$ and large $P_T$ approximation $M_{*T}\simeq P_T$,
we find the following approximate form for the $P_T$ distribution:
\begin{equation}
E_{M_*}\left.\frac{dN_{M_*}}{d^3P}\right|_{\eta=0}\sim 
{\rm e}^{-P_T / T_{\rm eff}^*}\, ,
\label{M*pt_distribution}
\end{equation}
where $T_{\rm eff}^*$ is the effective temperature for the preformed state defined by
\begin{equation}
T_{\rm eff}^*=T_{\rm reco}\ {\rm e}^{\eta^{}_T}=T_{\rm reco} \, \sqrt{\frac{1+v^{}_T}{1-v^{}_T}}\label{eff_temp}
\; .
\end{equation} 
The recombination temperature $T_{\rm reco}$ is identified with
the hadronization temperature in the original ReCo model.
The multiplicative factor $e^{\eta_T}>1$ reflects the effect of the nonzero transverse flow of the quarks,
which blue-shifts the inverse slope parameter of the preformed state distribution
from $T_{\rm reco}$ to $T_{\rm eff}^*$.
For example, this factor $T_{\rm eff}^*/T_{\rm reco}$ amounts to  $\sim $1.7 for $v_T=0.5$ and $2$ for $v_T=0.6$.

\subsubsection{Transverse momentum distributions of photons and mesons}

Given the distribution of the preformed states, let us discuss the
photon distribution eq.~(\ref{photon_distribution}) at mid-rapidity $k_L=0$.
We can perform the integration over $\Phi$ in eq.~(\ref{photon_distribution}) with the $\delta$-function
in the photon distribution (\ref{photon_CM_dist}) in the laboratory frame,
\begin{align}
  E_\gamma  \frac{dn^\gamma}{d^2k_T dk_L}
  &=  \frac{M_*}{4\pi k_0}\delta(
  k E_* - k_T P_T \cos(\Phi-\phi) - k_0 M_*)
  ,
\end{align}
where $\Phi$ ($\phi$) is the azimuthal angle of the preformed state (photon) from the reaction plane,
as shown in Fig.~{\ref{fig:decay}.
Then we obtain
\begin{align}
  \left .
  E_\gamma  \frac{dN_\gamma}{d^2 k_{T} dk_L}
  \right |_{k_L=0}  &=
\int_{-P_{L\rm max}}^{P_{L\rm max}} dP_L \int_{P_{T\rm min}}^{P_{T\rm max}} d P_T \, 
\frac{dN_*}{d^2P_T dP_L}\,
\frac{1}{4\pi k_0}
 \frac{2M_*}{k_T |\sin \theta|}
 \, ,
\label{PT_integral}
\end{align}
where 
\begin{align}
  \cos \theta =\cos(\Phi-\phi)=
  \frac{k E_{M*}  -k_0 M_*}{k_T P_T} 
  \, .
  \label{eq:theta}
\end{align}
The integration ranges of the longitudinal and transverse momenta,
$\pm P_{L\rm max}$ and $P_{T\rm min,max}$, of the preformed state are determined by decay kinematics (See Appendix A).

\begin{figure}[t] 
\begin{center}
  \includegraphics[width=0.4\textwidth,bb=0 0 600 600]{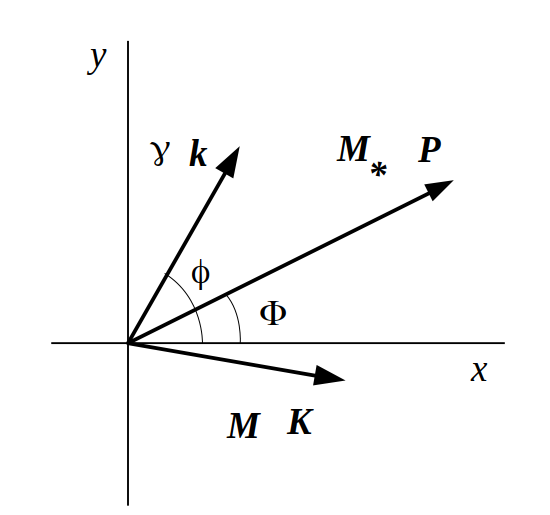}
\hspace{0.05\textwidth}
\includegraphics[width=0.35\textwidth,bb=0 0 450 300]{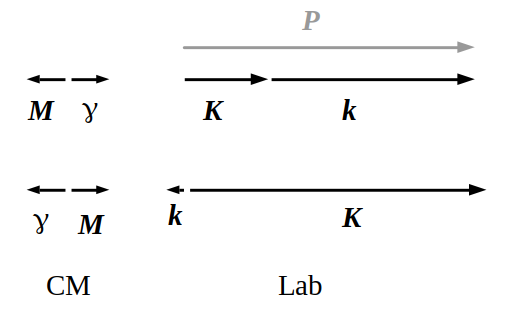}
\end{center}
\caption{Left: Kinematics of the photon emission from the preformed state, $\P \to \k + \K$.
  Right: Momentum boost from the CM to the laboratory frame
  for the photon momentum parallel (top) and anti-parallel (bottom) to $\P$.
}\vspace{-2mm}
 \label{fig:decay}
\end{figure}

We derive here approximate formulas valid for large photon momentum $k_T \gg M_*$.
The distribution of the preformed states (\ref{M*pt_distribution}) is a steeply-falling
function of $P_T$ for large $P_T \gg T_{\rm eff}^*$,
and therefore the dominant contribution to the integral comes from the lower end of the $P_T$ integration
with $P_L \simeq 0$.
In other words,
it comes from the configurations in which $\P$ is almost parallel to ${\bm k}$, 
{\it i.e.}, $\cos \theta \sim 1$ and $P_L \simeq 0$.
In this configuration,
the momentum $k_T$ is simply related to the CM-frame momentum $k_0$ by the transverse boost along $P_T = M_* \sinh y_T^*$ via
\begin{align}
  k_T=k_0 \cosh y_T^* + k_0 \sinh y_T^*  \sim 2k_0\frac{P_T}{M_*}
  =(1-\frac{M^2}{M_*^2}) P_T
\label{mom-shift}
\end{align}
as $\cosh y_T^* \sim \sinh y_T^*=P_T/M_*$.
The momentum deficit is carried by the accompanying meson emitted in the opposite direction in the
$M_*$-rest frame: $K_T= -k_0 \cosh y_T^* + \sqrt{M^2+k_0^2} \sinh y_T^*
\sim
(M^2/M_*^2)P_T ,
$
and $k_T + K_T = P_T$ holds as it should (Fig.~\ref{fig:decay} top-right). 

Since the number of the radiated photons is proportional to that
of preformed states, the photon distribution should behave as
(See Appendix A for more details)
\begin{align}
\left .  E_\gamma \frac{dN_\gamma}{d^2 k_T dk_L}
\right |_{k_L=0} \sim
\exp \left ( -\frac{P_{T}}{T^*_{\rm eff}(M_*)} \right ) 
\sim
       \exp \left ( -\frac{k_{T}}{T^\gamma_{\rm eff}(M_*)} \right )
\end{align}
with
\begin{align}
         T^\gamma_{\rm eff}(M_*)  = \left ( 1-\frac{M^2}{M_*^2} \right )
         T^*_{\rm eff}      \,  .
         \label{gamma_T}
\end{align}
Thus we find that the effective temperature of the photon $T^\gamma_{\rm eff}$ becomes lower than $T_{\rm eff}^*$.

On the other hand, the main contributions for the meson production by the radiative ReCo model come
from the configuration in which the meson momentum is parallel to $P_T$ in the $M_*$-rest frame
(See the bottom-right panel in Fig.~\ref{fig:decay}), 
the same calculation yields
\begin{align}
  K_T = k_0 \cosh y_T^* + \sqrt{M^2+k_0^2} \sinh y_T^* \sim P_T,
\end{align}
and the accompanied photon has nearly zero energy and momentum $k_T \sim 0$ due to its massless nature.
Accordingly, we find
\begin{align}
   \left . E_M \frac{dN_M}{d^2K_TdK_L}\right\vert_{K_L=0}
   \sim
   \exp \left(-\frac{K_T}{T_{\rm eff}^{\rm meson}(M_*)}\right)
\end{align}
with
\begin{align}
   T_{\rm eff}^{\rm meson}(M_*)=T_{\rm eff}^*
   \,   .
\label{eq:meson_T}
\end{align}
The meson has the same effective temperature as the preformed state, up to some corrections.

The radiative hadronization predicts that there is an ordering in the effective temperatures of
photons, mesons, and preformed states: 
\begin{equation}
  T_{\rm eff}^\gamma(M_*) < T_{\rm eff}^{\rm meson}(M_*)
  \sim T_{\rm  eff}^* \, .
  \label{eq:ordering}
\end{equation}
Note that the thermal exponential form of the photon and meson distributions reflects the shape of the parton distributions
and thus the origin of higher effective temperatures of photons and mesons than the hadronization temperature $T_{\rm reco}$
can be attributed to the partonic collective flow built up during the QGP evolution.

\subsubsection{$v_2$ of photons and mesons}

Within the same approximations, we can evaluate the elliptic flow
coefficient for the photons, $v_2^\gamma(k_T)$, defined by 
\begin{equation}
  v_2^\gamma(k_T)\equiv \frac
  {{\displaystyle
      \int d\phi \cos 2\phi
      \left(\left. k \frac{dN_\gamma}{d^2k_Tdk_L }\right|_{k_L=0} \right)}}
  {{\displaystyle
      \int d\phi
      \left(\left. k \frac{dN_\gamma}{d^2k_T dk_L }\right|_{k_L=0}\right)}}\, .
  \label{v2_gamma}
\end{equation}
The distribution of the preformed state is computed with the original ReCo model,
\begin{equation}
\frac{dN_*}{d^2P_T dP_L}=
   \overline{ \frac{dN_*}{d^2P_T dP_L}}\,
   \Big(1+2v_2^*(P_T)\cos 2\Phi \Big)\, ,
\label{v2_preformed}
\end{equation}
where 
$\overline{dN_*/d^2 P_T dP_L}$ is $\Phi$-independent part of the spectrum and the nonzero elliptic flow $v_2^*(P_T)$ is inherited from the quark/anti-quark elliptic flow.
Then, in place of eq.~(\ref{PT_integral}), the integration over $\Phi$ with the $\delta$-function
yields
\begin{align}
  \left .
  k \frac{dN_\gamma}{d^2 k_{T} dk_L}
  \right |_{k_L=0}  &=
\int_{-P_{L\rm max}}^{P_{L\rm max}} dP_L \int_{P_{T\rm min}}^{P_{T\rm max}} d P_T \,
 \overline{\frac{dN_*}{d^2P_T dP_L}}\,
 (1+2v_2^*(P_T) \cos 2 \phi \cos 2 \theta)\frac{1}{4\pi k_0}
 \frac{2M_*}{k_T |\sin \theta|}
,
\label{gamma_dist_2}
\end{align}
where we have used $\sum_{i=\pm}\cos 2\Phi_i = 2 \cos 2\phi \cos 2\theta$
with $\Phi_\pm = \phi \pm \theta$.
We insert eq.~(\ref{gamma_dist_2}) into the definition (\ref{v2_gamma}),
and evaluate the momentum integral approximately with its threshold value near $P_T \sim P_{T\rm min}$ and $P_L \sim 0$,
where we can put $\cos 2\theta \sim 1$.
Thus we find that, after the $\phi$ integration, the elliptic flow coefficient is unchanged
but its momentum argument is replaced with $k_T = (1-M^2/M_*^2)P_T$~:
\begin{equation}
  v_2^\gamma(k_T) \sim v_2^*\left(\frac{k_T}{1-M^2/M_*^2}\right)
  \, .
\label{photon_v2_scaling}
\end{equation}
In the same manner, we find the coefficient for the meson distribution in radiative ReCo model as 
\begin{equation}
v_2^{\rm meson}(K_T)\sim v_2^*\left(K_T \right)\, .
\label{meson_v2_scaling}
\end{equation} 
We emphasize here that the elliptic flow coefficients $v_2$ of the photons and mesons are both given by $v_2^*$ of the preformed states, 
which approximately satisfies the CQN scaling, $v_2^*(P_T) \sim 2 v_2^q(P_T/2)$~(see eq.~\eqref{eq:CQNscaling}).

We have shown that the effective temperature and the elliptic flow coefficient of 
the meson distribution are the same as those of the preformed-state distribution, 
while for the photon distribution these parameters are estimated in the similar manner but with a simple momentum shift \eqref{mom-shift}.
This means that even if the contributions of the radiative hadronization are added,
not only does the meson elliptic flow still follow the CQN scaling,
but also the photon elliptic flow does so approximately.

\section{Numerical results}

For our numerical study we employ a two-dimensional (2D) model, neglecting the longitudinal momentum  of the preformed state $M_*$ ($P_L=0$),
since we have seen that the $P_L$-integration plays only a minor role in the modification of the photon and meson distributions from that of the preformed state.

The photon distribution of the 2D radiative ReCo model reads 
\begin{align}
  \left .
  k \frac{d N_\gamma}{d^2k_T dk_L} \right |_{k_L=0}
  &=\kappa \int_{P_{T\rm min}}^{P_{T\rm max}} dP_T  \,
  \sum_{i=\pm} \frac{dN_{M_*}}{d^2P_TdP_L}(P_T, \Phi_i,0)
  \frac{1}{2\pi} \frac{M_*}{k_T  |\sin \theta |}
  \, ,
  \label{2d_photon_distribution}
\end{align}
where 
$\Phi_\pm=\phi\pm \theta$ with $\theta$ defined in eq.~\eqref{eq:theta}.
We generate the distribution of the preformed states of mass $M_*$,
using the original ReCo model \cite{Fries:2003kq,Fries:2003vb}.
But we change the recombination temperature $T_{\rm reco}$ to 155 MeV, which is within the range of the pseudo-critical temperature obtained
in lattice QCD calculations \cite{Borsanyi:2010cj,Bazavov:2011nk}.
Accordingly the transverse flow at hadronization is set to $v_T=0.6$ to reproduce the $p_T$ distribution of the mesons,
and the freeze-out time $\tau$ and the fireball radius $\rho_0$ are also adjusted adequately (see Table \ref{tab:ParamSet}).
We set $M_* = 2M_{ud}$ for the preformed state based on the constituent quark model picture.

\begin{table}
  \begin{tabular}{cccccccccc}
  \hline
&  $T_{\rm reco}$~/MeV & ~ $v_T$ ~  & ~$\tau$~/fm ~ & ~ $\rho_0$~/fm ~ &~ $\gamma_{u,d}$~ & ~$\gamma_{\bar u,\bar d}$~ &  $M_{ud}$~/MeV  & $p_0$~/GeV & $a$ \\
  \hline
  RHIC &   155 & 0.6  &  8.0  & 12.5 & 1  & 0.9 &  260  &  1.0 & 2.5\\
  LHC  &   155 & 0.65 &  15.0 & 20.0 & 1  & 1   &  260  &  1.1 & 2.5\\ 
\hline
\end{tabular}
  \caption{
    Model parameters.
  }
\label{tab:ParamSet}
\end{table} 

\subsection{Model characteristics}
First let us numerically test the characteristics of particle distributions of our radiative ReCo model,
which was discussed in the previous section.
Note that in this subsection we set the normalization factor $\kappa=1$ of the radiative ReCo model to study the model characteristics,
while in the next subsections we will adjust the parameter $\kappa$ so that the model reproduces the observed photon distributions.

\subsubsection{Transverse momentum spectrum}
We show $p_T$ distributions of the pions and photons produced by the radiative ReCo model at $\sqrt{s_{NN}}=200$ GeV
in Fig.~\ref{fig:model_char_pion} (left), along with the distribution of the preformed states.
At a given momentum, yields of the pions and photons are much lower than that of the preformed state because
each of them shares a fraction of the momentum of the preformed state,
and therefore their distributions are shifted to the lower $p_T$ region.
But their $p_T$ slopes are similar to each other in the relevant momentum region.

We define the effective temperature $T_{\rm eff}$ here
by fitting each of the $p_T$ distributions with an exponential function $\exp(-p_T/T_{\rm eff})$
in the momentum range $2 < p_T < 5$ GeV.\footnote{%
 The slope parameter $T_{\rm eff}$ is centrality-independent in this model
 since the quark distribution \eqref{phase_space_dist}
 depends on collision centrality only by the weak modulation of the transverse flow \eqref{etaT_phi}
 and by the overall factor of the hot-zone size $f(r,\phi)$.
}~
In Fig.~\ref{fig:model_char_pion} (right) we check the $M_*$ dependence of these effective temperatures,
$T_{\rm eff}^{\pi}$ of pions and $T_{\rm eff}^\gamma$ of photons produced in the radiative ReCo model.
They follow the ordering eq.~\eqref{eq:ordering} derived in the previous section.
$T_{\rm eff}^\pi = T_{\rm eff}^*$ is predicted by eq.~\eqref{eq:meson_T}, and 
the difference between $T_{\rm eff}^{\pi}$ and $T^*_{\rm eff}$ may be understood
as a correction due to subleading $p_T$-dependence of the distributions.
On the other hand, $T_{\rm eff}^\gamma$ of photons is lower than $T^*_{\rm eff}$ and
approaching it with increasing $M_*$, as predicted by eq.~\eqref{gamma_T}. 

Our model restricts the invariant mass of the preformed state to $M^*= 2 M_{ud}=520$~MeV for pion production,
but in more general treatments, the preformed states with $M^* \ge 2M_{ud}$ may well contribute to the pion production.
However, from Fig.~\ref{fig:model_char_pion} (right), the slope parameter $T_{\rm eff}$ of the photons and pions from the radiative ReCo model
seem rather insensitive to the $M_*$ value, as long as $M_*$ is much larger than the meson mass $M$.

\begin{figure}[th]
 \centering
 \includegraphics[width=7.5cm]{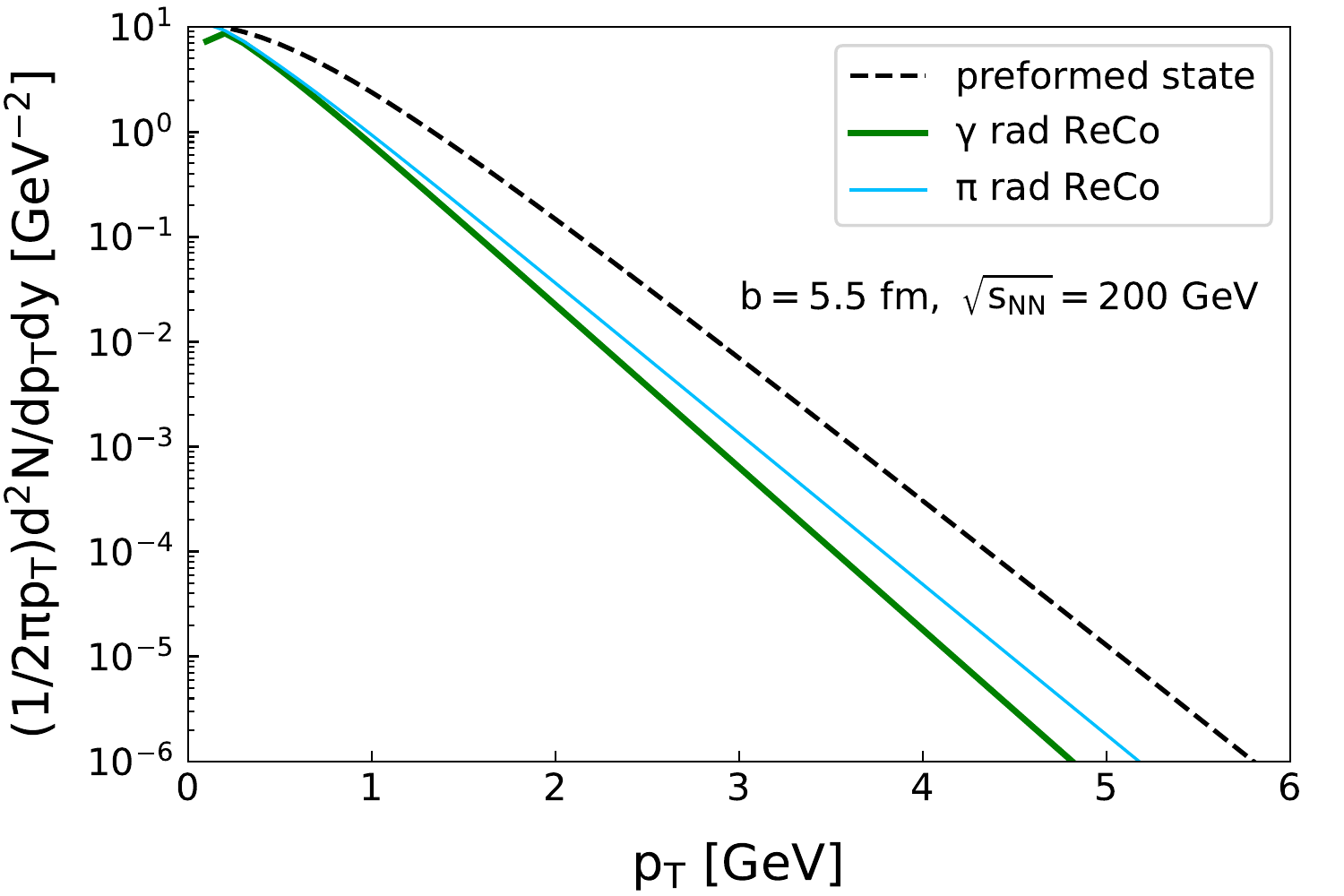}
 \hfil
\includegraphics[width=7.5cm]{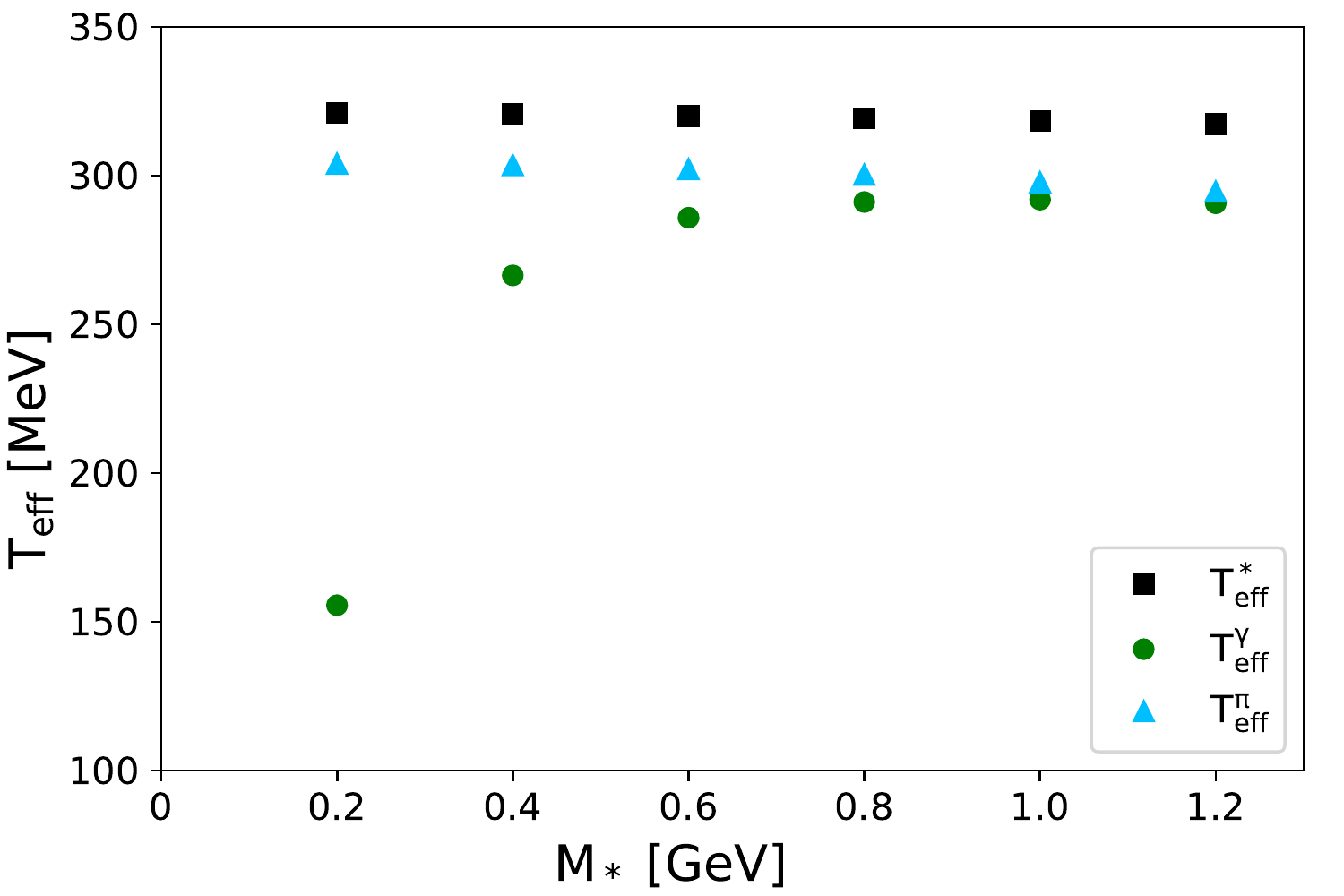}
 \caption{%
   Left: Comparison of transverse momentum distributions $d^2N/(2\pi p_T dp_T dy)$
   of the photons (green bold solid), pions (blue thin solid), and preformed state (black dashed) in the radiative ReCo model
   ($M_* = 2M_{ud}$, $\kappa=1$ with $b=5.5$ fm at $\sqrt{s_{NN}}=200$ GeV).
   Right: 
$M^*$-dependence of the slope parameters of the photons $T_{\rm eff}^\gamma$ (green circle),
   of pions $T_{\rm eff}^\pi$ (blue triangle), and of preformed states $T_{\rm eff}^*$ (black square).
   The parameter $T$ is determined by fitting the function $\propto e^{-p_T/T}$ 
   to the $p_T$ distributions in the momentum range $2 < p_T < 5$ GeV.
   \label{fig:model_char_pion} } 
\end{figure}

We show the results for kaon production with $M_* = M_s + M_{ud}$ and $M_s = 460$ MeV in Fig.~\ref{fig:model_char_kaon},
where the particle yields become smaller at $p_T \lesssim 2$ GeV than the pion case due to the mass effect. 
Moreover, the $p_T$ slope of the photon distribution is much steeper than those of the kaon and preformed state distributions, 
because, unlike the pion mass, the kaon mass $M_K=495$~MeV is comparable to $M_*=720$~MeV here (see eq.~\eqref{gamma_T}).
We see that the photon yield associated with radiative hadronization of the kaons is small compared with that of the pions,
and we neglect it in this work.

\begin{figure}[th]
 \centering
 \includegraphics[width=7.5cm]{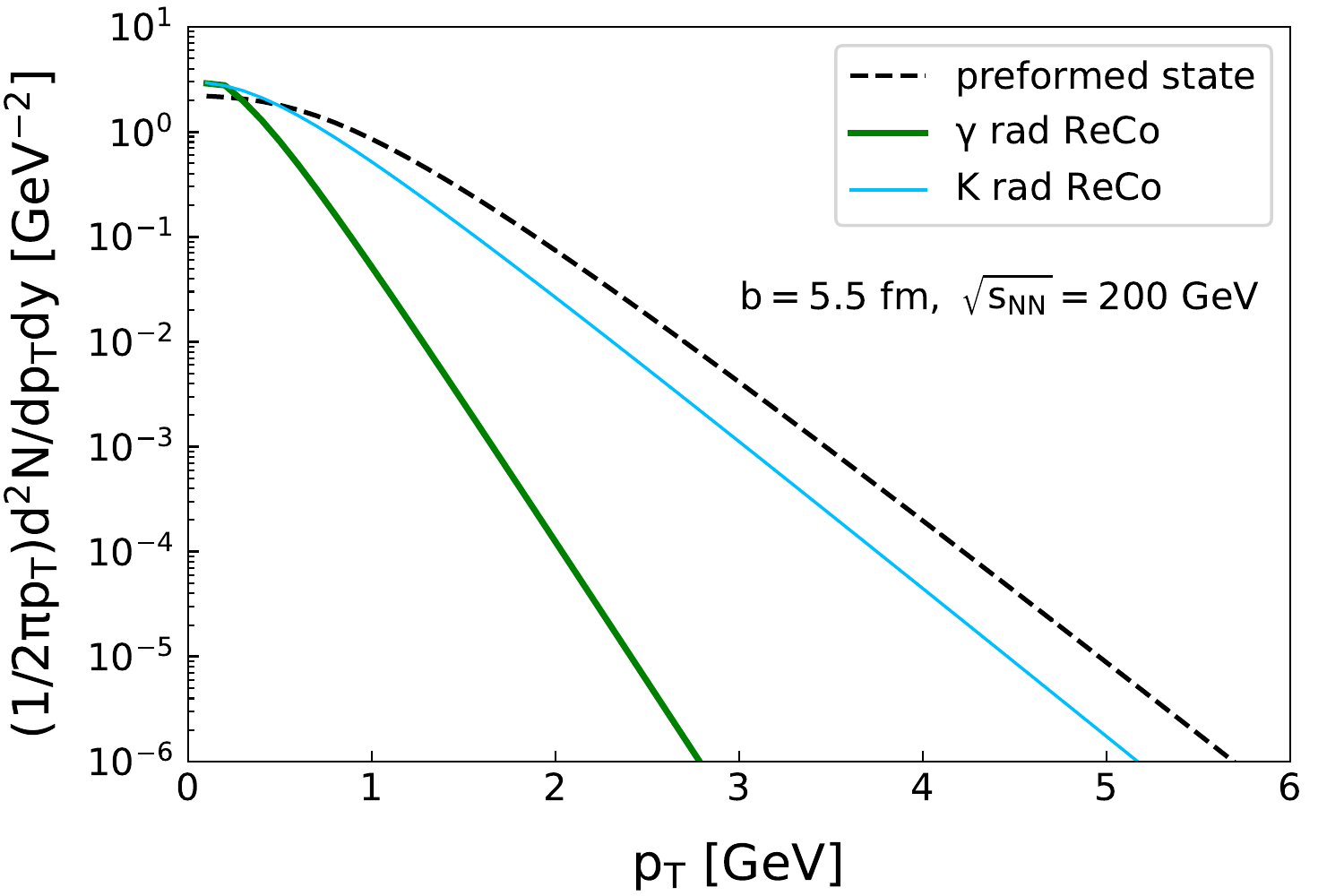}
 \caption{%
    The same plot as in Fig.~\ref{fig:model_char_pion} (right),
  but associated with kaon production ($M_s =460$ MeV).
  \label{fig:model_char_kaon}
 }
\end{figure}

\begin{figure}[th]
 \centering
 \includegraphics[width=7.5cm]{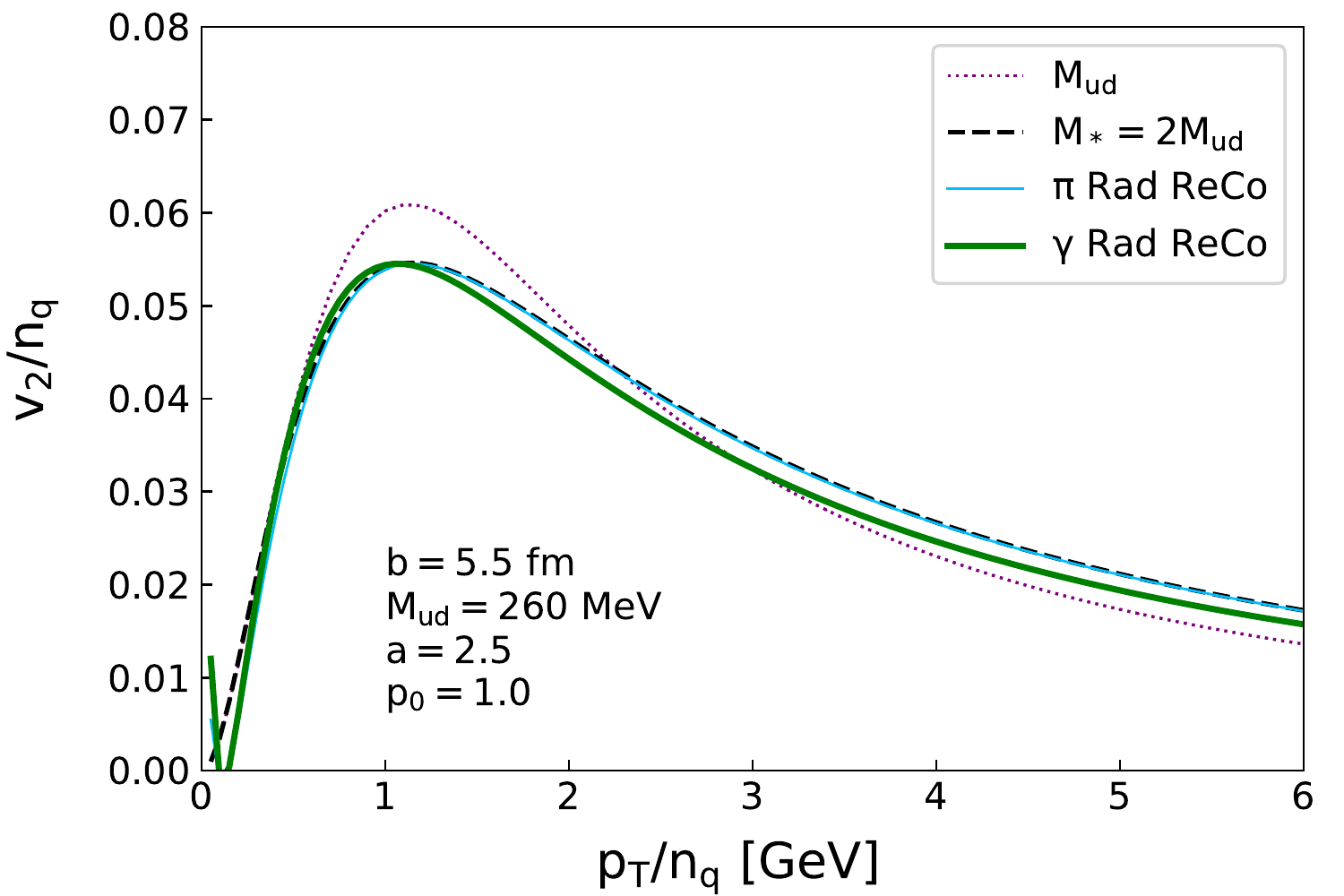}
 \caption{%
   Rescaled elliptic flow coefficients $v_2(p_T/n_q)/n_q$
   of the pions (blue thin solid), photons (green thick solid),
   and preformed state (black dashed) at $b=5.5$ fm with $n_q=2$.
   The flow coefficient $v_2(p_T)$ of the quarks (purple dotted)
   is shown for comparison.
   Parameters are the same as in Fig.~\ref{fig:model_char_pion}.  
   \label{fig:check_QNS_v2_rhic}
 }
\end{figure}

\subsubsection{Elliptic flow $v_2$}

We introduce an azimuthal angle dependence of the quark/antiquark flow $v_T(p_T)$
by the modulation amplitude $h(p_T)$ of the transverse flow rapidity $\eta_T(\phi; p_T)$ as in eq.~(\ref{etaT_phi}).
The magnitude of $h(p_T)$ is determined by the aspect ratio of the initial collision zone.
In Fig.~\ref{fig:check_QNS_v2_rhic} we show the result of the elliptic flow coefficients $v_2$
of the pions (cyan thin) and photons (green bold) as well as that of the preformed states (black dashed),
after dividing ed by the constituent quark number $n_q=2$.
The quark/antiquark elliptic flow coefficient $v_2^q$ is also shown (purple dotted) for comparison.

We find that all these $v_2$ have the same magnitude,
which is taken over from the quark/antiquark flow through the radiative recombination.
We also note that the flow coefficient $v_2^\pi$ of the pions exactly follows
that of the preformed state $v_2^*(p_T)$ at $p_T \gtrsim 1$ GeV,
while the photon $v_2^\gamma(p_T)$ lies slightly below them.
Indeed, we have confirmed that,
when plotted in the shifted momentum $\bar p_T=p_T/(1-M^2/M_*^2)$,
the photon $v_2^\gamma(p_T)$ curve overlaps with that of the preformed states at larger $p_T$,
as shown in eq.~\eqref{photon_v2_scaling}.

\subsection{RHIC}

Next we study the contributions of radiative hadronization to the photon production in Au+Au collisions at $\sqrt{s_{NN}}=200$ GeV.
In order to make a comparison of the photon yield to the direct photon data at RHIC \cite{Adler:2003qi}, 
we include the thermal photon contributions evaluated with a three-dimensional viscous hydrodynamic model with the kinetic freezeout temperature $T_{\rm fo}=116$ MeV \cite{Miyachi-Nonaka}
(see Appendix B for a concise model description).
For calculating the particle distribtions in the original and radiative ReCo models,
we assign the impact parameter $b=3.0, 5.5, 7.5$ and $9.0$ fm in our model for the centrality classes, 0--10\%, 10--20\%, 20--30\% and 30--40\% of the collision events, respectively, based on the Glauber model estimate \cite{PHENIX:2003iij},
while the hydrodynamic model treatment includes the event-by-event fluctuations
in selecting the centrality classes.

In Fig.~\ref{fig:centrality_rhic} we compare the $p_T$ distributions of $\pi^0$ obtained
by the ReCo (black solid) and radiative ReCo (cyan dashed) models to PHENIX data at different centralities \cite{Adler:2003qi}.
The parameter $\kappa=0.2$ of the radiative ReCo model is determined so that
the sum of the photons from radiative hadronization, the thermal photons and the prompt photons
reproduces the observed photon yield (see Fig.~\ref{fig:photon_centrality_rhic} below).
We are reassured here that the original ReCo model reproduces the pion $p_T$ distributions for different centrality classes, in the $p_T$ range from 2 to 4 GeV,
where the quark recombination is regarded as the dominant hadronization mechanism.
Outside this region, other production mechanisms, hydrodynamic processes at the lower $p_T$ and parton fragmentations at the higher $p_T$, are important.
In contrast, the contribution from the radiative ReCo model takes only a small fraction of the pion yield (less than 10~\% of the original ReCo model) at a given momentum $p_T$,
and it does not obstruct the success of the original ReCo model in describing meson production in this momentum region \cite{Fries:2003kq,Fries:2003vb}.

\begin{figure}[th]
 \centering
 \includegraphics[width=10cm]{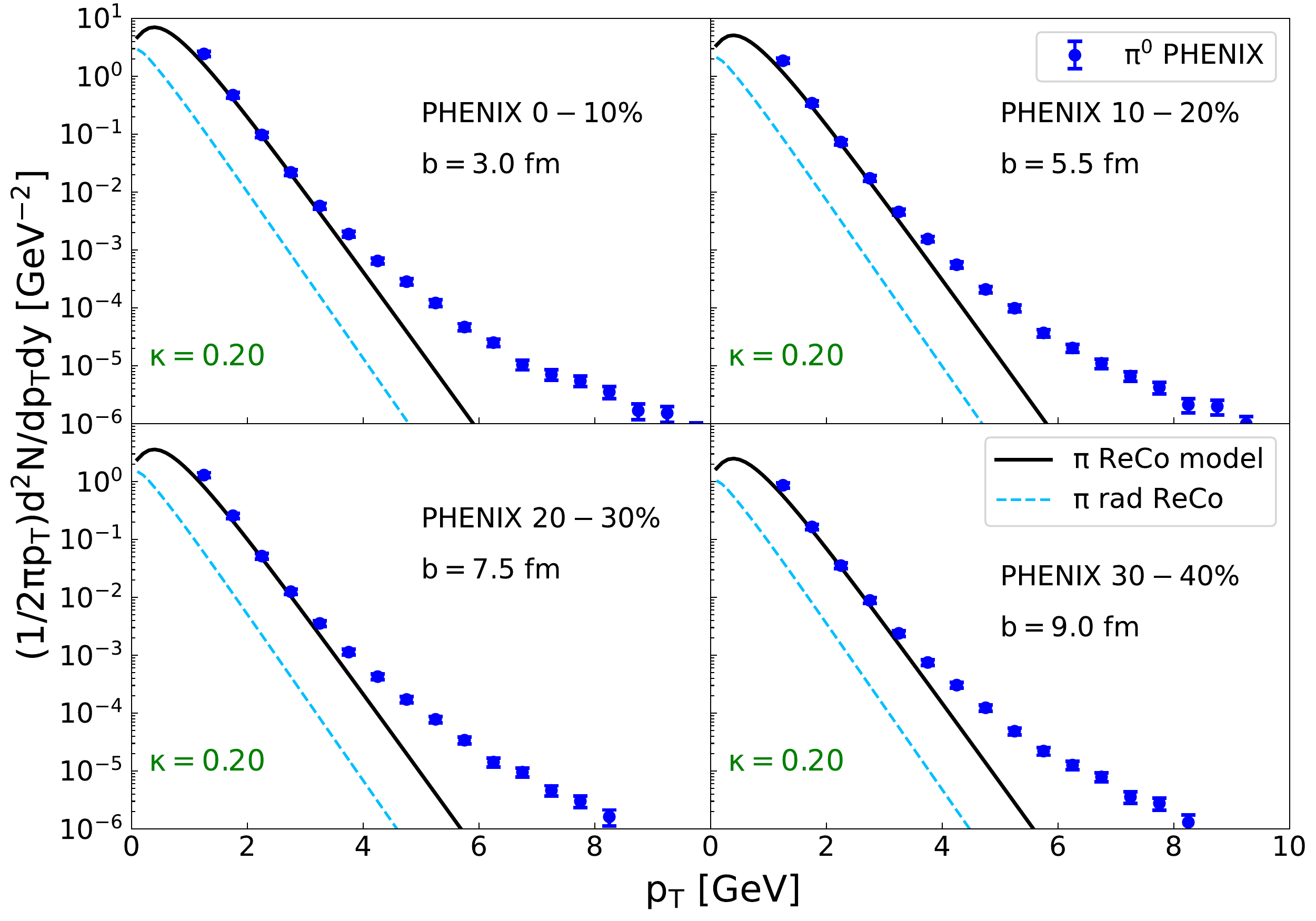}
 \caption{Transverse momentum distributions of $\pi^0$
   computed with ReCo model (black solid) and radiative ReCo model (cyan dashed)
   for impact parameter $b=3$, 5.5, 7.5 and 9 fm in Au+Au  collisions
   at $\sqrt{s_{NN}}=200$ GeV.
   The $\pi^0$ data set of 0--10 \%, 10--20 \%, 20--30 \% and 30--40 \%  centrality classes
   is adopted from \cite{Adler:2003qi}.
   \label{fig:centrality_rhic} } 
\end{figure}

\begin{figure}[th]
  \centering
  \includegraphics[width=10cm]{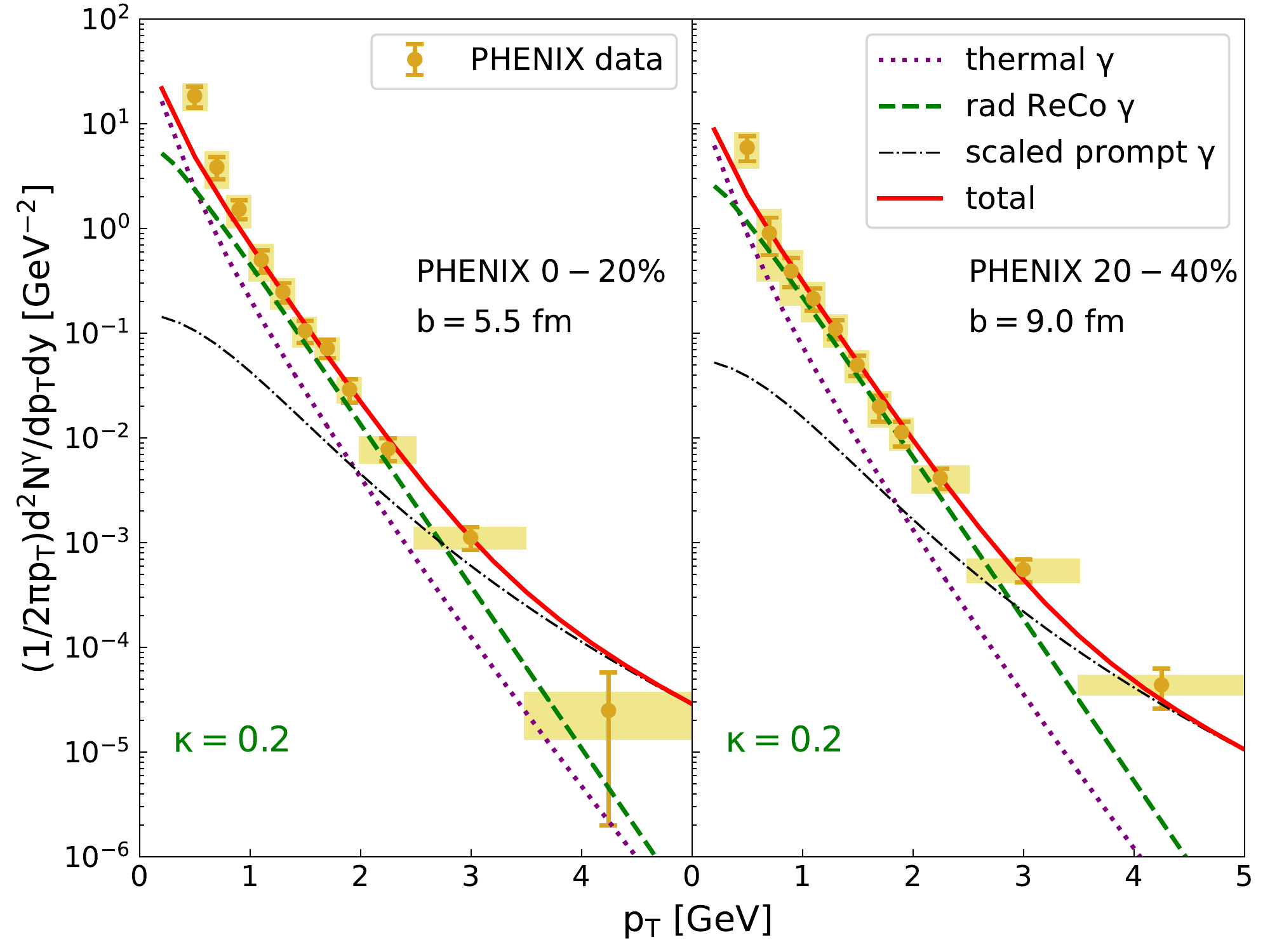}
 \caption{
 Transverse momentum distributions of direct photons 
 computed with the radiative ReCo model (green dashed) for impact parameter $b=5.5$ (left) and $9$ fm (right)
 in Au+Au collisions at $\sqrt{s_{NN}}=200$ GeV.
The thermal photon distribution obtained by a viscous hydrodynamic model (purple dotted),
 rescaled prompt photons (black dot-dashed),  and their sum (red solid) are also shown.
 The data are adopted from \cite{Adare:2014fwh}, and the parameter $\kappa=0.2$ is determined to fit the data. 
\label{fig:photon_centrality_rhic}
}
\end{figure}

\begin{figure}[th]
 \centering
 \includegraphics[width=10cm]{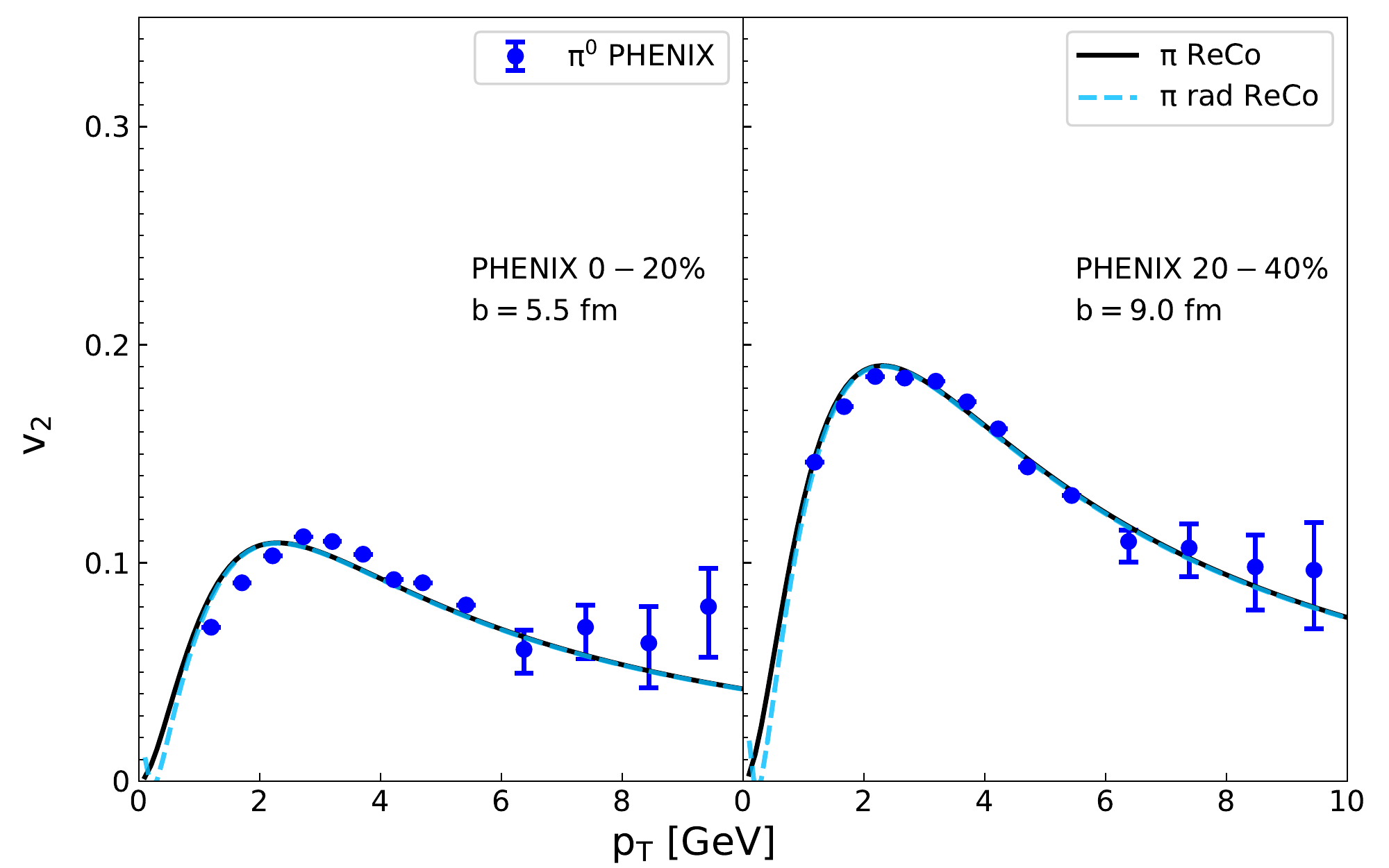}
 \caption{
   Pion elliptic flow coefficient $v_2$ from the ReCo model (black solid)
   and the radiative ReCo model (blue dashed) as a function of $P_T$
   for $b=5.5$ fm (left) and $b=9.0$ fm (right) in Au+Au collisions at $\sqrt{s_{NN}}=200$ GeV. 
   Data for $\pi^0$ (blue circles) are taken from PHENIX \cite{Adare:2013wop}.
   \label{fig:v2_rhic}
}
\end{figure}

\begin{figure}[th]
 \centering
 \includegraphics[width=10cm]{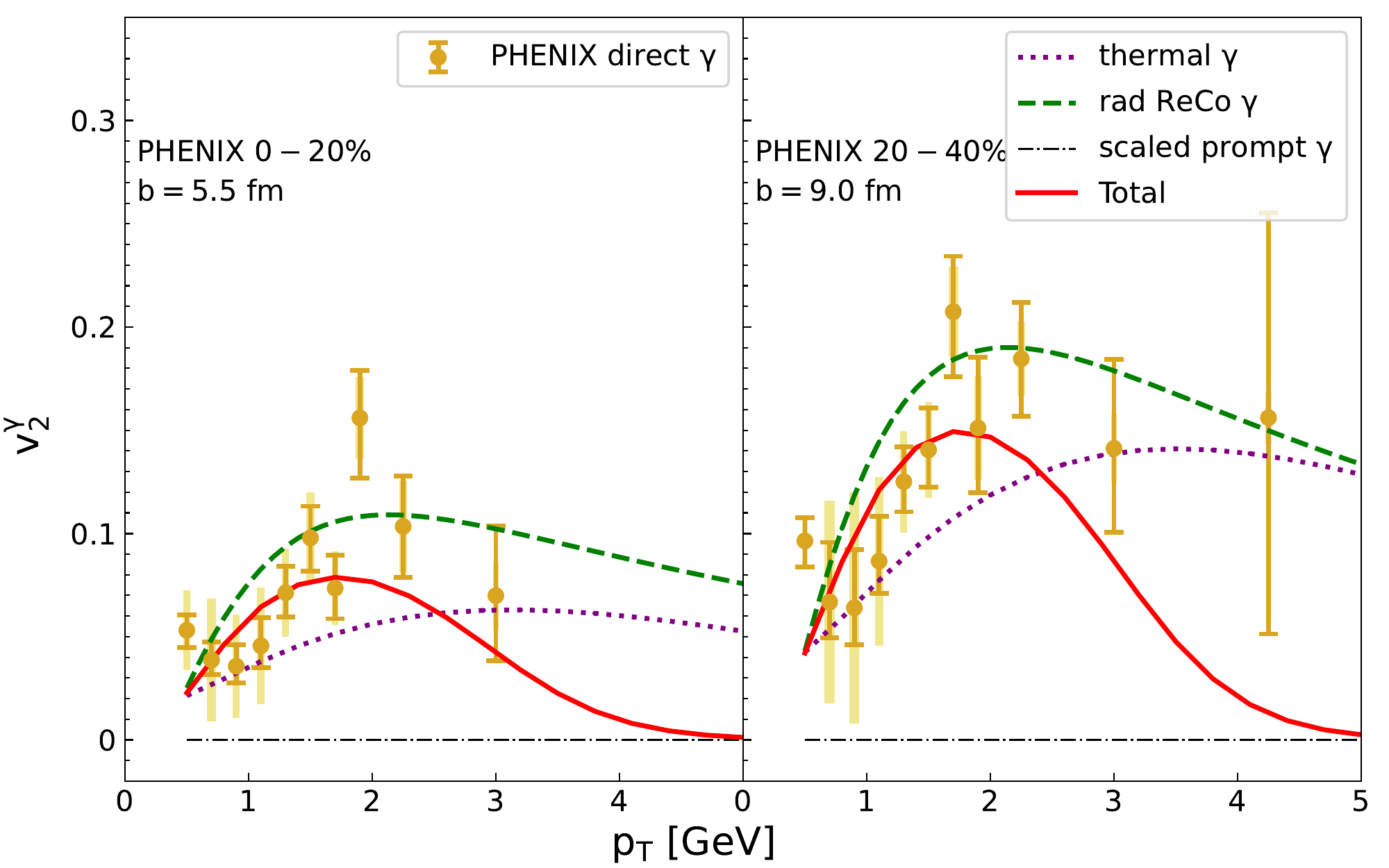}
 \caption{
   Elliptic flow coefficient $v_2^\gamma$ of the direct photons (red solid) for impact parameter
   $b=5.5$ (left) and 9 fm (right) in Au+Au collisions at $\sqrt{s_{NN}}=200$ GeV.
 The $v_2^\gamma$ of the photons from a viscous hydrodynamic model 
 and $v_2^\gamma$ of the photons from radiative ReCo model are shown in purple dotted and green dashed curves, respectively.
The normalization $\kappa=0.2$ for the radiative ReCo model is adopted.
   Data for direct photons (blue solid stars) is adopted from \cite{Adare:2015lcd}.
   \label{fig:v2_photon_rhic}
}
\end{figure}

Next in Fig.~\ref{fig:photon_centrality_rhic}
we show the $p_T$ distributions of the photons emitted in radiative pion production (green dashed)
for $b=5.5$ fm (left panel) and $b=9.0$ fm (right panel),
together with those of the thermal photons (purple dotted) and the prompt photons (black dot-dashed),
and the total (red solid).
For thermal photon production, we adopt the thermal photon rates of QGP in \cite{Arnold:2001ms} and
that of the hadronic phase in \cite{Holt:2015cda,Turbide:2003si,Heffernan:2014mla},
and integrate these rates over the evolution profile obtained by a 3D viscous hydrodynamic simulation (See Appendix B for a concise model description).
Our estimate of the thermal photon contribution is consistent with other model studies \cite{Paquet:2015lta}.
Regarding prompt photon production in AA collisions, which dominates the total photon distribution at higher $p_T$,
we use the empirical fit of the photon distribution in pp collisions, $a_1 (1+p_T^2/a_2)^{a_3}$ ($a_{1,2,3}$ are constants),
scaled with the number of nucleon collisions for AA collisions, 
as is done by PHENIX \cite{Adare:2014fwh}.
The experimental data of the direct photons in the 0--20 \% (for $b=5.5$ fm) and 20--40 \% (for $b=9.0$ fm) centrality classes
are taken from PHENIX \cite{Adare:2014fwh}\footnote{%
  A new data analysis was published in Ref.~\cite{PHENIX:2022rsx} and
  is consistent with that in Ref.~\cite{Adare:2014fwh}.
}.

We set the normalization of the radiative ReCo model to $\kappa=0.2$ so that
the sum of the three photon contributions, thermal radiation, radiative hadronization, and prompot production,
reproduces the observed photon yield for $p_T < 3$ GeV.
Indeed, the photon $p_T$ distributions for two centrality classes $0-20$ \% and $20-40$ \% are reproduced fairly well with the same normalization $\kappa=0.2$.
We notice that the photon yield from the radiative ReCo model is estimated to be several times larger than that from the thermal radiation and that the $p_T$ slope of the resultant photon distribution is mostly determined by the contribution from the radiative ReCo model
for $1.5<p_T<3 $ GeV and is consistent with the data.
Notice that the prompt photon contributions become non-negligible only for the higher $p_T \gtrsim 3$ GeV.

Next we study the elliptic flows of the pion and photon distributions.
We present in Fig.~\ref{fig:v2_rhic}
the elliptic flow coefficient $v_2(p_T)$ of the pion distribution obtained by the ReCo model (solid line) as well as that of the radiative ReCo model (dashed line), separately,
despite knowing that the latter's contribution to pion production is very small.
We find that the two curves overlap for $p_T \gtrsim 1$ GeV.

Concerning the photon $v_2^\gamma$ in our model, we recall that there are three different sources of the photons, {\it i.e.}, the thermal radiation, the radiative hadronization, and the prompt production.
In Fig.~\ref{fig:v2_photon_rhic} we show the flow coefficient $v_2^\gamma(p_T)$ (red solid) of the total photon distribution as well as those of the thermal photon contribution (purple dotted) and of the radiative ReCo model contribution (green dashed), separately.
In addition, we presumed the prompt photons have no collective flow (black dot-dashed).
The thermal photons have a nonzero $v_2^\gamma$ but its value is systematically below the observed values \cite{Adare:2014fwh}.  
On the other hand, the photons from the radiative ReCo model have $v_2^\gamma$ as large as the pion $v_2$, and its $p_T$ dependence is almost the same as that of the pions.
Since the photon yield of the radiative ReCo model is estimated to be several times larger than the thermal photon yield, the resultant $v_2^\gamma(p_T)$ of the total photons is close to that of the radiative ReCo model component and is consistent with the data for $p_T \lesssim  2$ GeV, albeit with a large uncertainty. At larger $p_T$, the prompt photons dominate and the flow is suppressed.

\subsection{LHC}

We perform the same analysis for photon production in heavy-ion collisions at the LHC energy, $\sqrt{s_{NN}}=2.76$ TeV.
In this calculation we use the same recombination temperature $T_{\rm reco}=155$ MeV,
but with a stronger transverse flow $v_T=0.65$ of the quarks at the hadronization.
We also assume full equilibration of $\bar{u}$ and $\bar{d}$ quarks and set the fugacity factors $\gamma_{\bar{u}}=\gamma_{\bar{d}}=1$.
See Table~\ref{tab:ParamSet} for other parameters.
We adopt here the following correspondence between the impact parameter $b$ and the centrality classes,
$b=4.0$ fm for 0--10~\%, $6.0$ fm for 10--20~\%,
$7.8$ fm for 20--30~\% and $9.2$ fm for 30--40~\% centrality class \cite{ALICE:2013hur}.

In Fig.~\ref{fig:central-pt-lhc} shown are the $p_T$ distributions of the pions
for several centrality classes.
The original ReCo model (black solid) describes the $p_T$ spectra of the pions in the region $1<p_T<3$ GeV at different centralities%
\footnote{Note that hydrodynamic processes and jet fragmentations contribute to the pion production at $p_T < 1$ GeV and $p_T > 4$ GeV,
respectively.}.
With the normalization $\kappa=0.05$, which we determine so as to reproduce the photon yield below,
the radiative ReCo model (cyan dashed) gives only a few percent of the pion yields in the relevant momentum region.
Hence the success of the original ReCo model is unaltered by the contribution of the radiative ReCo model.

\begin{figure}[th]
\centering
\includegraphics[width=10cm]{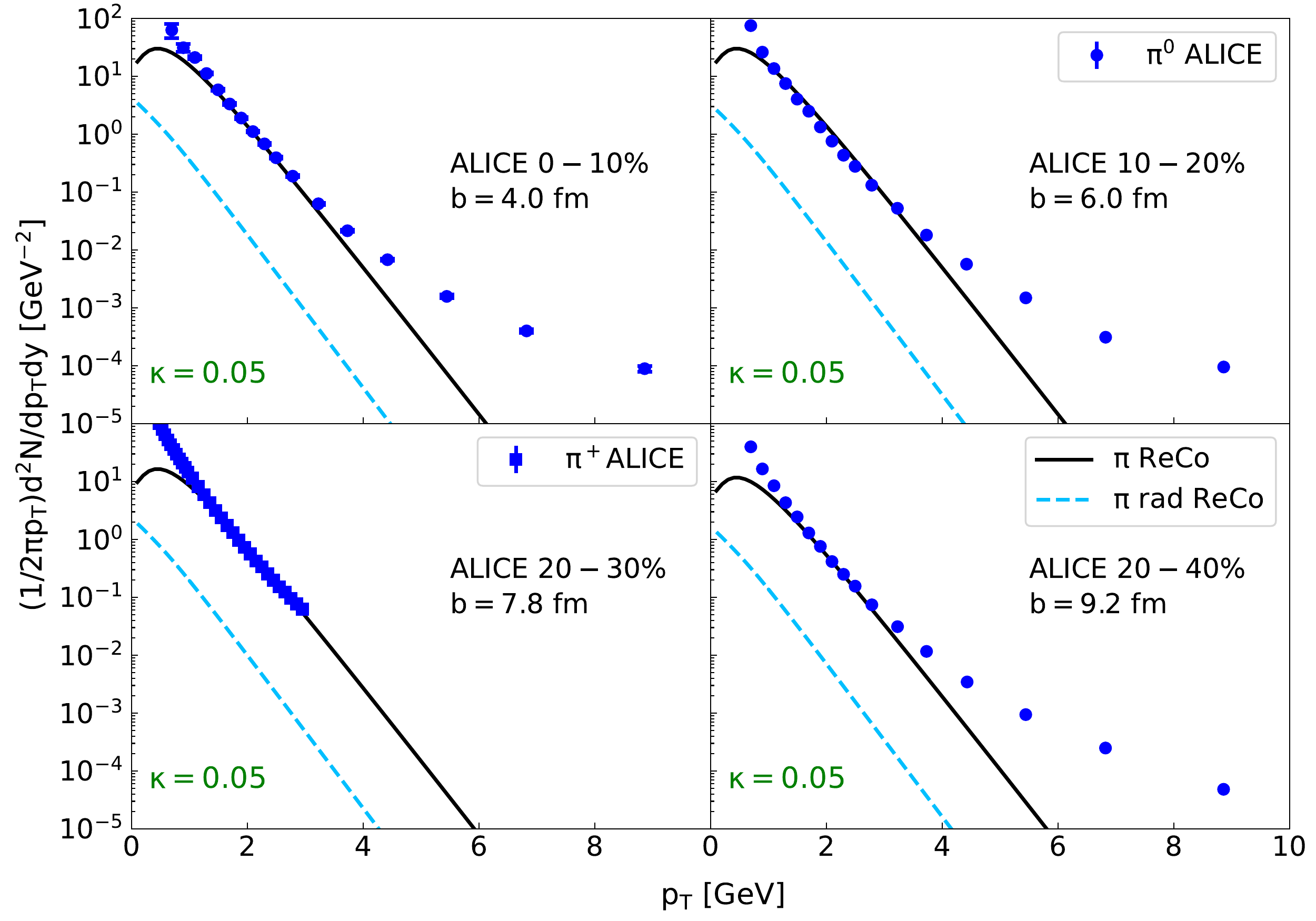}
 \caption{
 Centrality dependence of transverse momentum distributions of $\pi$ 
 in $\sqrt{s_{NN}}=2.76$ TeV Pb+Pb collisions at the LHC.
 Data points for $\pi^0$ in the
 0--10 \%, 10--20 \% and 20--40 \%  centrality classes  \cite{Abelev:2014ypa} 
 and  $\pi^+$ in the 20--30 \% centrality class \cite{Abelev:2013vea}
 are adopted from ALICE.  
 \label{fig:central-pt-lhc}
} 
\end{figure}

In Fig.~\ref{fig:pt_lhc} 
we compare the photon $p_T$ distribution of our model at $b=6.0$ (9.2) fm
with the experimental data \cite{Abelev:2013vea} at 0--20 (20--40) \% centrality.
Unlike in the RHIC case, the thermal photon yield (purple dotted) is not far off the observed data in the momentum region $1 < p_T < 2$ GeV,
and we can reasonably fit the data in the region $p_T \lesssim 2$ GeV
by adding the photons from the radiative ReCo model (green dashed) with the normalization factor $\kappa = 0.05$, both in central and mid-central collisions.
Since prompt photon data in pp collisions is unavailable in this low $p_T$ region, and theory calculations for $p_T \lesssim 2$ GeV have a large ambiguity,
we decided here to use the same model of photon distribution $a_1 (1+p_T^2/a_2)^{a_3}$ as adopted by PHENIX \cite{Adare:2014fwh},
setting $a_2=4$ GeV$^2$ and
tuning the parameter $a_1=1.2\times 10^{-2}$ GeV$^{-2}$ and $a_3 =-2.7$ to fit the available pp-collision data around $p_T \sim 10$ GeV at $\sqrt{s}=8$~TeV \cite{ALICE:2018mjj}.
The parameter $a_2$ may be regarded as an infra-red cutoff of the prompt photon model.
We estimate the prompt photons in AA collisions by rescaling this model \cite{Adare:2014fwh}. 
We note here that the normalization parameter $\kappa$ is not very sensitive to the choice of $a_2$ as long as it is of order $O(1~\text{GeV}^2)$.
Although we obtained a reasonable fit of the data in this low $p_T$ region with $\kappa=0.05$,  we don't have a clear explanation for the decrease of the $\kappa$ value
from the RHIC case.
(We will come back to this point in the next section.
See also Appendix C for $\kappa$ dependence.)

\begin{figure}[th]
 \centering
 \includegraphics[width=10cm]{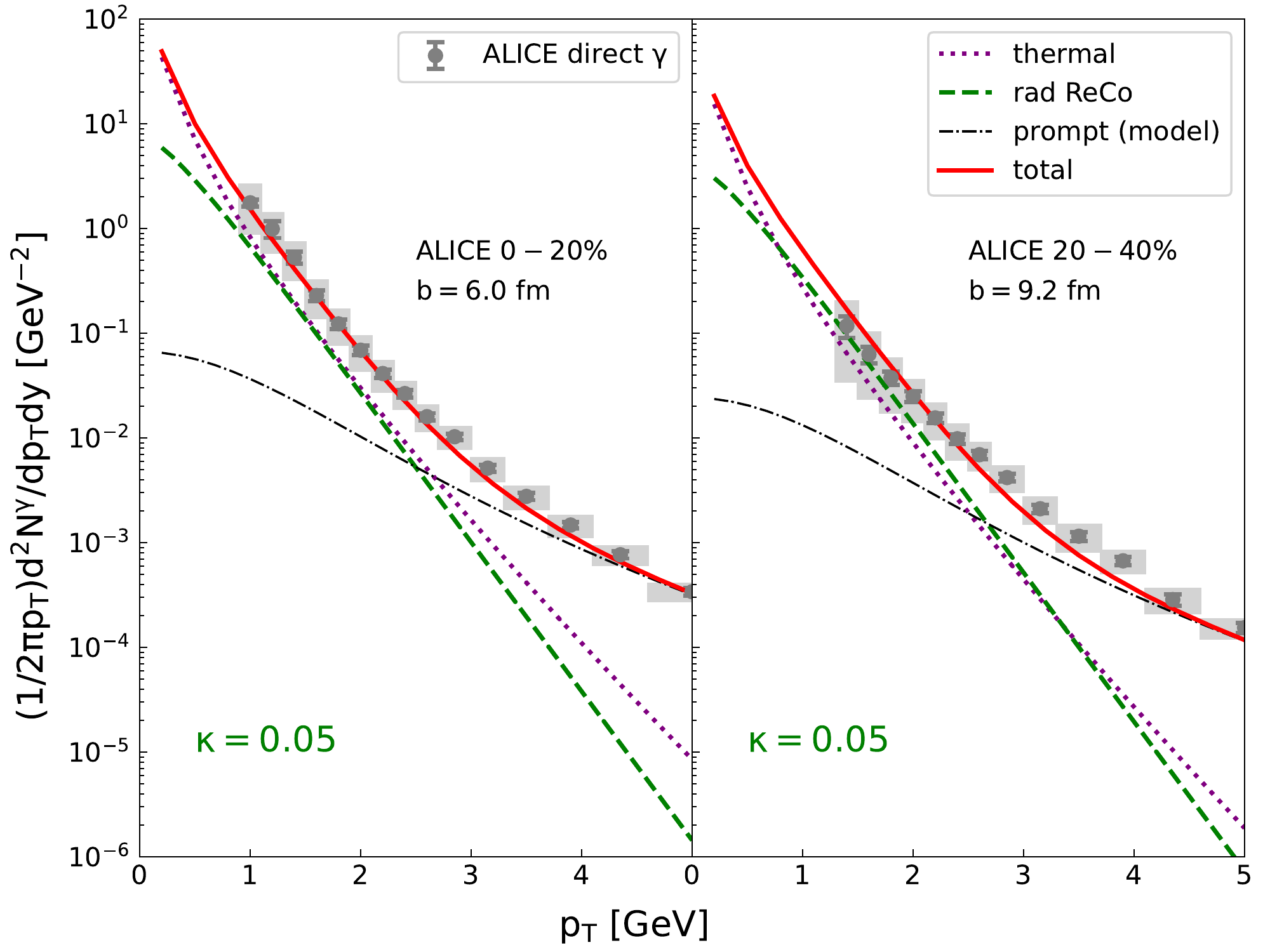}
 \caption{
   Transverse momentum distributions of direct photons 
 (red solid) for $b=6$ fm (left) and  $b=9.2$ fm (right). 
 The photon yields from the hydrodynamic model and the radiative ReCo model are shown
 in purple dotted and green dashed curves, respectively.
 The data in the corresponding centrality classes (gray circles) are adopted from \cite{Adam:2015lda}.
 \label{fig:pt_lhc} 
} 
\end{figure}

Next we study the elliptic flow coefficient $v_2$ of the pions and photons.
We use the parameters $p_0 = 1.1$ GeV and $a=2.5$ for the quark elliptic flow $v_2(p_T)$ (see eq.~\eqref{hpt-profile}).
Figure \ref{fig:pion_v2_lhc} shows the pion elliptic flow $v_2$ for $b=6.0$ (left) and 9.2 fm (right)
along with ALICE data \cite{Adam:2015lda}.
We find that the pion $v_2$ is well described by the original ReCo model
and the situation is unaltered by the addition of the radiative hadronization contribution.

Figure \ref{fig:v2_lhc} shows the photon elliptic flow coefficient $v_2^\gamma$ for $b=6.0$ (left) and 9.2 fm (right).
Since in $1 < p_T < 2$ GeV the thermal radiation and radiative hadronization contribute almost equally to the photon yield,
the elliptic flow coefficient $v_2^\gamma$ of the total photon yield becomes an average of the two sources and
lies just in between $v_2^\gamma$ of the thermal photon (purple dotted)  and $v_2^\gamma$ of radiative ReCo photons (green dashed) at lower $p_T$.
Indeed, this is consistent with the fact that the photon $v_2^\gamma$ is somewhat smaller than the pion $v_2$ in the observed data at the LHC energy.
At higher $p_T$, it is dominated by the prompt photon contribution, which we assume has zero elliptic flow.

\begin{figure}[th]
 \centering
 \includegraphics[width=10cm]{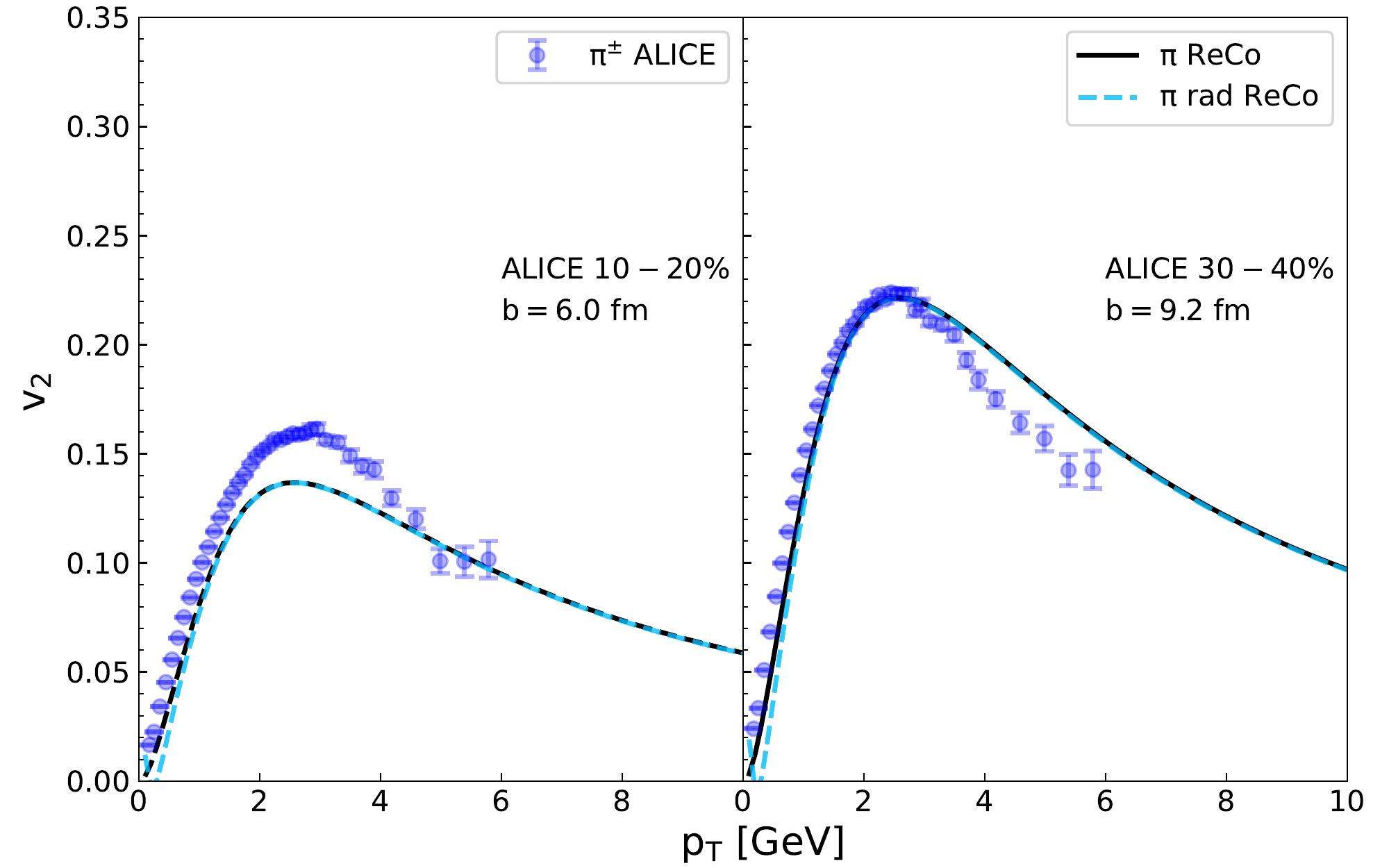}
 \caption{
   Elliptic flow $v_2$ of the pions of ReCo model (black solid) and of radiative ReCo model (cyan dashed) for $b=6.0$ (left) and 9.2 fm (right), 
corresponding to centralities, 0--20 \% and 20--40 \%, respectively.
Note that centralities of data (blue circles) are 10--20 \% (left) and 30--40 \% (right) \cite{ALICE:2014wao}, respectively.
   \label{fig:pion_v2_lhc}} 
\end{figure}

\begin{figure}[th]
 \centering
 \includegraphics[width=10cm]{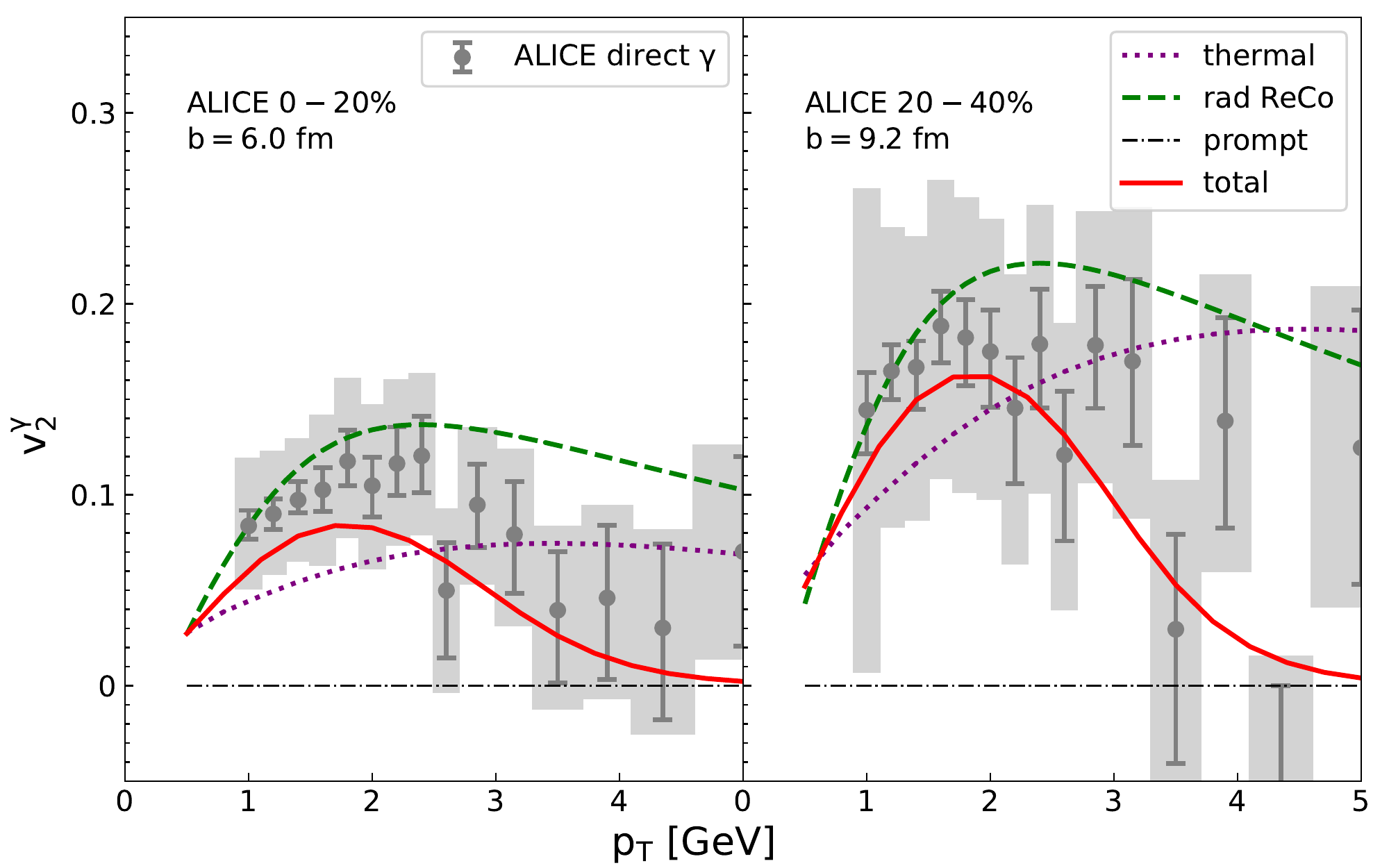}
 \caption{
Elliptic flow $v_2$ of direct photons (red solid) for $b=6.0$ (left) and 9.2 fm (right).
Data (gray circles) are adopted from \cite{Lohner:2012ct}. 
Other notations are the same as in Fig.~\ref{fig:v2_photon_rhic}.
   \label{fig:v2_lhc}} 
\end{figure}

\section{Summary and discussions}

In QED plasma physics, radiative recombination is one of the fundamental processes, which occur when a plasma is de-ionized.
In analogy with it,
we have proposed here ``radiative hadronization'' \cite{Fujii:2017nbv}
as an additional photon source which had not been taken seriously so far in heavy-ion collision physics.

In this paper, we have embodied the radiative hadronization process as a two-step model, that is, picking up a preformed state of quarks,
which we describe by the original ReCo model \cite{Fries:2003kq,Fries:2003vb}, and its radiative decay to a photon and a hadron.
We call this model the ``radiative ReCo'' model.
Those particles produced by the radiative ReCo model inherit collective flows from the quark fluid.
We can thus expect that those photons fill up the deficit of the direct photon sources to reproduce the observed yield
and that they give rise to the strong collective flow of photons.
We derived approximate formulas for the effective temperatures, $T_{\rm eff}^\pi$ and~$T_{\rm eff}^\gamma$,
and elliptic flows of pions and photons, $v_2^{\pi}$ and $v_2^\gamma$, in this model,
and found that their elliptic flow coefficients still satisfy the CQN scaling relation approximately~(Fig.~\ref{fig:check_QNS_v2_rhic}).

For numerical study of direct photons, we considered three kinds of photon sources, prompt photons, thermal photons and photons from the radiative hadronization.
Photon production from the radiative hadronization was evaluated with the two-dimensional radiative ReCo model, while thermal photon yield was
obtained by integrating  the rate formulas over hydrodynamic-model event profiles and a simple parametrization was used for prompt photon yield.

We have shown that the radiative ReCo model can account for a significant fraction of the total photon yield, while it produces only a small fraction of the total pion yield,
to which the original ReCo gives a dominant contribution at around $ p_T \sim  2$~GeV.
Indeed, by adding the photon contribution from the radiative ReCo model with the overall factor $\kappa=0.2$ to the thermal and prompt photon contributions,
we can fit {\it simultaneously} the large direct photon yield and the strong elliptic flow observed at RHIC.
For the LHC data, the same procedure resulted in a reasonable fit of the $p_T$ distribution and the elliptic flow $v_2^\gamma(p_T)$ of photons in $ 1 < p_T < 4$~GeV,
but with the smaller value of $\kappa=0.05$.
The apparent reason for this decrease of $\kappa$ is the fact that the estimated thermal photon yield is not far off the observed direct photon yield compared to the RHIC case.
The origin of this change and the large values of $\kappa$ as well is unclear at present, but we
expect that it would contain important information about hadronization processes, which clearly
deserves further investigations.


Hadronization from QGP is a non-perturbative and non-equilibrium process in high-energy heavy-ion collisions.
The empirical success of the recombination models seems to suggest that complexity of hadronization is involved to some extent in these models.
The overall factor $\kappa$ of the radiative ReCo model may be interpreted as a ratio of the radiative recombination to the original recombination.
The obtained values $\kappa =0.2$ for RHIC and 0.05 for the LHC data are indeed small, but still larger than the naive expectation of $O(\alpha_{\rm em})=1/137$.
(In Appendix C, as a reference, we studied the sensitivity of the photon $p_T$ spectrum and elliptic flow $v_2^\gamma$ to the $\kappa$ parameter for the RHIC data.)
These $\kappa$ values should reflect various dynamical effects in hadronization.
For example, there would be quark recombination process with gluon radiation which turns an octet quark pair into a singlet one, which is
parametrically of order $O(\alpha_s)$. But, of course, the colored gluons cannot come out from the collision zone.
On the other hand, the photons are penetrating from the zone.

Concerning the preformed states in our model,
it is admittedly unclear whether we can assign any resonance interpretation to them and we intentionally avoid such an interpretation
due to the lack of our understanding of the hadronization mechanism.
In this work, we have assumed that the invariant mass distribution of the preformed state, or a quark pair in a simple term, is peaked at $M_*=2m$,
and replaced it with a delta-function. If we relax this restriction, we will have
\begin{equation}
  k \left.\frac{dN_\gamma}{d^2k_T^\gamma dk_L}\right\vert_{k_L=0}
  \sim
  \int_{2m}^\infty dM_*\, \varrho(M_*)\, \# \, \exp \left(-\frac{k_T^\gamma}{T_{\rm eff}^\gamma(M_*)}\right)
\end{equation}
with $T_{\rm eff}^\gamma(M_*)$ being the same as in eq.~(\ref{gamma_T}).
The $M_*$ distribution $\varrho(M_*)$ should modify our estimate of the photon yield,
but part of the modification may be effectively absorbed in the adjustable overall factor $\kappa$.

Incidentally, the pion production in radiative hadronization implies that the preformed state with spin 1 decays to the pseudo-scalar state with a photon emission,
but our radiative ReCo model does not specify the spin quantum number of the preformed state.
The parameter $\kappa$ may include this kind of uncertainties, too.

We have studied direct photon production in an {\it ad-hoc} way by simply adding three sources, prompt photons, thermal photons and photons from the radiative hadronization of pions,
evaluated with different models.
The radiative hadronization of heavier hadrons give much smaller contribution to the photon yield at $p_T \sim 2$ GeV (Fig.~\ref{fig:model_char_kaon}).
But there are other possible photon sources. Just to name a few examples,
jet fragmentation produces photons at relatively higher $p_T$, and an initial pre-equilibrium evolution called the glasma stage also contributes to the photon yield thought its elliptic flow will be small \cite{Berges:2017eom,Monnai:2019vup}.
In this sense, we should keep in mind that our estimate for $\kappa$ may be regarded as  an upper bound for the radiative hadronization contribution.

Photon production is a unique penetrating probe, along with the dilepton production. 
The so-called ``direct photon puzzle'' has been an important challenge in understanding the evolution of heavy-ion collisions, QGP formation and its decay.
A unified model for describing the collision events is strongly desired
and we believe that the radiative hadronization occupies a crucial position in understanding the photon production in heavy ion collision physics.

  \begin{acknowledgments}

    This work was supported in part by JSPS KAKENHI Grant Numbers, JP16K05343 and JP21K03568 (HF), and
    JP20H00156, JP20H00581 and JP17K05438 (CN), and JP19K03836 (KI).

  \end{acknowledgments}
  

\appendix
\section{Momentum distribution of the decay particle}


We consider generically the decay process of a particle of mass $M$ to two particles of masses $m_1$ and $m_2$: $M \to m_1 + m_2$.
The momentum distribution of the produced particle $m$ (being $m_1$ or $m_2$)
with transverse momentum $k_T$ at mid-rapidity $k_L=0$
is expressed as
\begin{align}
  \left .
  \epsilon \frac{dN_m}{d^2 k_{T} dk_L}
  \right |_{k_L=0}  &=
\int \frac{d^2 P_T dP_L}{E}\,
E \frac{dN_M}{d^2P_T dP_L}\,
\epsilon  \frac{dn_m}{d^2 k_{T} dk_L}
\, ,
\label{eq:kt-distribution}
\end{align}
where
$\epsilon = \sqrt{m^2 + \k^2}$, and
$E =\sqrt{M^2+\P^2}$ with $\P=(\P_T, P_L)$ the three-momentum of the particle $M$.
We assume the produced particle distribution is isotropic 
in the $M$-rest frame,
\begin{align}
\epsilon_{\CM}  \frac{dn_{m}}{d^3 k_{\CM}}=
\frac{1}{4\pi k_0}\delta (\epsilon_{\CM} -\epsilon_{0})
\end{align}
where $\epsilon_{0}$ is the energy of the produced particle
in the $M$-rest frame:
\begin{align}
  \epsilon_{0}=\sqrt{m^2+k_0^2}\; , \quad
  k_0=\sqrt{[M^2 -(m_1+m_2)^2][M^2 -(m_1-m_2)^2]\, }/2M
  \; .
\end{align}
When one of the produced particles is a photon: $M \to m + \gamma$,
we set one of $m_{1,2}$ to zero.

In order to evaluate the integral \eqref{eq:kt-distribution},
it is convenient to parametrize the momentum of the particle $M$
with two successive boosts from the $M$-rest frame,
a longitudinal boost with rapidity parameter $y_L$ and then a transverse one with $y_T$:\footnote{%
  Note that $y_L$ is different from the standard longitudinal rapidity $y$ which is defined by $\tanh y= P_L /E$.}
\begin{align}
    (M, 0,0,0) &\xrightarrow{y_L}
  (M \cosh y_L, 0,0,M \sinh y_L)
    \notag \\
    &\xrightarrow{y_T}
    (M_{L} \cosh y_T, M_{L} \sinh y_T \cos \Phi,
    M_{L} \sinh y_T \sin \Phi,M\sinh y_L) \equiv (E,\P_T, P_L).
\end{align}
  Here $M_L\equiv \sqrt{M^2+P_L^2}=M\cosh y_L$,
  and $\Phi$ is the azimuthal angle of the transverse momentum ${\bm P}_T$.
  Let us take the frame in which $\Phi=0$ so that the transverse boost is along the positive $x$.
By the two successive boosts,  the momentum
  $k_{\CM}^\mu \equiv (\epsilon_{0},
   k_{{\CM}x}, k_{{\CM}y}, k_{{\CM}L})$
   of the particle $m$ in the $M$-rest frame is transformed as
   \begin{align}
  (\epsilon_{0}, k_{{\CM}x}, k_{{\CM}y}, k_{{\CM}L})
&    \xrightarrow{y_L}
    (\epsilon', k_{{\CM}x}, k_{{\CM}y}, k'_L)
   \xrightarrow{y_T}
    (\epsilon, k_{x}, k_{y}=k_{{\CM}y}, k_L=k'_L)
\end{align}
with 
\begin{align}
  \left \{ \begin{array}{l}
  \epsilon' ~ = \epsilon_{0}\cosh y_L + k_{{\CM}L}\sinh y_L \, ,
  \\
      k'_L = k_{{\CM}L}\cosh y_L + \epsilon_{0} \sinh y_L \, ,
  \end{array}
  \right .
  \qquad
  \left \{
  \begin{array}{l}
   \epsilon ~ \;  =\epsilon' \cosh y_T + k_{{\CM}x}\sinh y_T \,  ,
\\ 
 k_x = k_{{\CM}x}\cosh y_T + \epsilon' \sinh y_T \, .
  \end{array}
  \right .
\end{align}
Here a prime $(')$ denotes the quantities in the frame boosted by $y_L$ from the CM frame.

Now let us determine the range of the momentum integration in \eqref{eq:kt-distribution}.
First, the condition $k_L=0$ fixes the parameter $y_L$ as $\tan y_L = - k_{{\CM}L}/\epsilon_0$.
This $|y_L|$ takes its maximum $\tan y_{L\rm max} = k_{0}/\epsilon_0$
when the particle momentum $\k_{\CM}$ is along the beam axis, $k_{{\CM}L}= \pm k_0$, in the $M$-rest frame.
Thus the integration range of $P_L$ of the particle $M$ is found to be $-P_{L\rm max} < p_L < P_{L\rm max}$ with
\begin{align}
  P_{L{\rm max}} \equiv M\sinh y_{L\rm max} = M\frac{k_0}{m}\, .
\end{align}
For photon production $m=0$, it becomes $P_{L{\rm max}}=\infty$.

Next, consider the transverse boost with $y_T$, which maps the transverse momentum $\k_{{\CM}T}$
on the circle in the primed frame
to $\k_T$ on an ellipsoid in the laboratory frame (Fig.~\ref{fig:boost} (top)):
\begin{align}
  k_{{\CM}T}^2 =k_{{\CM}x}^2 + k_{{\CM}y}^2 =
   \left (  \frac{k_x - \epsilon' \sinh y_T}{ \cosh y_T} \right )^2 
  + k_y^2 
  \, .
\end{align}
This $y_T$-boost transformation is concisely expressed as
\begin{align}
  y_m = y'_m + y_T
\end{align}
in terms of the transverse rapidities of the particle, defined as
$y'_m=\tfrac{1}{2}\ln (\epsilon'+k_{{\CM}x}) / (\epsilon'-k_{{\CM}x})$ and
$y_m=\tfrac{1}{2}\ln (\epsilon+k_{x})/(\epsilon-k_{x})$
in the primed and laboratory frames, respectively.
For a given $k_T$ in the laboratory frame, the boost rapidity $y_T$ takes its smallest value
when $\k_{{\CM}T}$ is parallel to $\k_T$ as shown in Fig.~\ref{fig:boost} (bottom).
Indeed, the magnitude of the boost rapidity lies within the range
\begin{align}
y_{T\rm min} \equiv |y_{m_{\rm MAX}}  -y'_{m_{\rm MAX}} | < y_T < y_{T\rm max} \equiv   y_{m_{\rm MAX}}  + y'_{m_{\rm MAX}}
\end{align}
with
\begin{align}
  y_{m_{\rm MAX}}  = \tfrac{1}{2}\ln \frac{ \epsilon+k_{T}      }{\epsilon-k_{T}}
  \, , \qquad
  y'_{m_{\rm MAX}} = \tfrac{1}{2}\ln \frac{\epsilon'+k_{{\CM}T}}{\epsilon'-k_{{\CM}T}}
  \, .
\end{align}
This determines the upper and lower limits of the integration over
$P_T=M_L\sinh y_T$ for given momenta $k_T$ and $k_{{\CM}T}$ of the particle.
For production of a photon $m_1=0$ at $k_L=0$,
the relations $\epsilon=k_T$ and $\epsilon'=k_{{\CM}T}$ hold
and the integration range becomes
\begin{align}
|\ln \frac{k_T}{k_{{\CM}T}}| \le y_T \le \infty
  \, .
\end{align}

\begin{figure}[t] 
\begin{center}
\includegraphics[width=0.7\hsize]{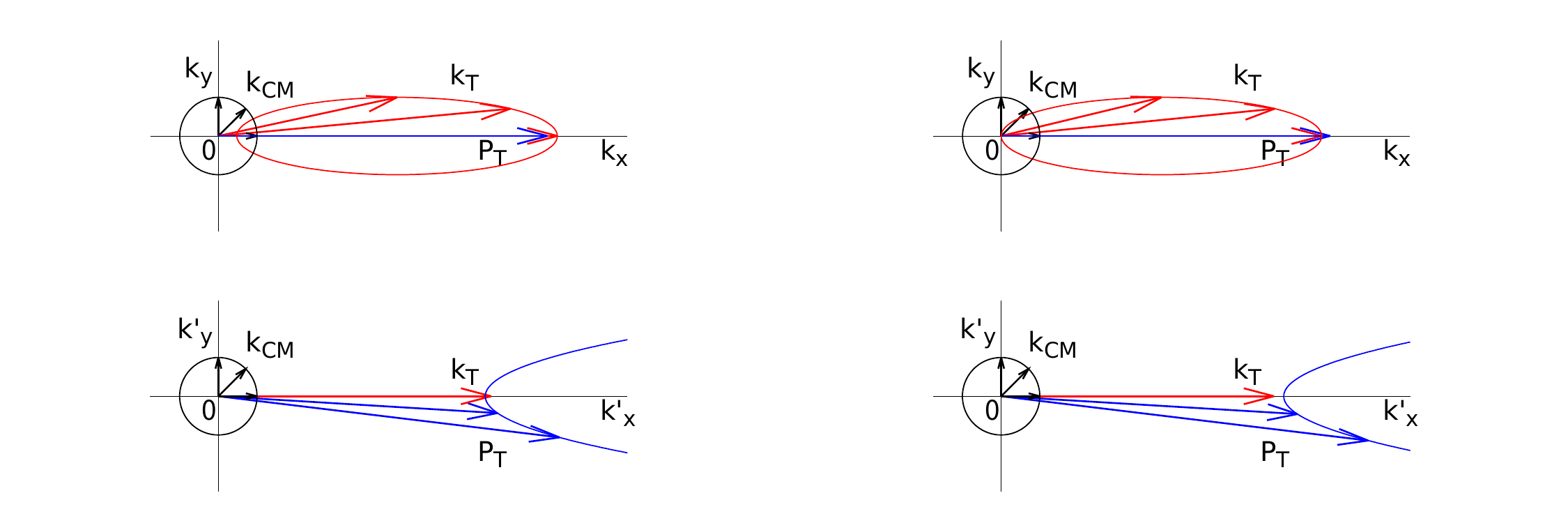}
\end{center}
\caption{
  The particle momentum $k_{{\CM}T}$ in the CM frame is boosted by the transverse rapidity $y_T$ of the parent particle $M$
  ($P=M\sinh y_T$) to form an ellipsoid in the laboratory frame ($k_{{\CM}L}=0$ for simplicity).
  With a strong enough boost a massive particle is made to move along the boost direction (top-left),
  but a massless photon moving in the opposite direction only becomes soft without changing its moving direction (top-right).
  On the other hand, for a fixed $k_T$,
  the momentum of its parent particle $P_T$ becomes smallest when $\P_T$ and $\k_{{\CM}T}$ are parallel:
  $P_{T{\rm min}} = k_T+k_{2T}$ with $k_{2T}$ being the momentum of the accompanying particle.
  The $k_{2T}$ becomes nearly zero for large-$k_T$ photon production (bottom-left),
  while it gets close to $(m^2/M^2)P_{T\rm min}$ for large-$k_T$ particle production (bottom-right).
  }
  \label{fig:boost}
\end{figure}

Now that we have determined the integration range of \eqref{eq:kt-distribution}
in terms of the rapidities $y_L$ and $y_T$ of the particle $M$,
we can write the formula \eqref{eq:kt-distribution}
more explicitly using these variables.
First, we restore the azimuthal angle $\Phi$ of the momentum $\P$,
and perform the $\Phi$ integration in \eqref{eq:kt-distribution}
using the $\delta$-function constraint.
In terms of the laboratory-frame momenta,
$\P=(P_T\cos\Phi, P_T \sin \Phi, P_L)$ and
$\k=(k_T \cos \phi, k_T \sin \phi, 0)$,
the constraint is written as 
\begin{align}
\delta(\epsilon_{\CM} - \epsilon_0)
&  =
 \delta(\frac{k \cdot  P}{M}  - \epsilon_0)
 =
  \frac{M}{k_T P_T } \delta(
  \cos(\Phi - \phi)  - \cos \theta)
  \; ,
\end{align}
where $\cos \theta$ is defined as
\begin{align}
  \cos \theta
 \equiv
    \frac{\epsilon E - \epsilon_0 M}{k_T P_T}
  =
  \frac{\epsilon \cosh y_T \cosh y_L - \epsilon_0}{k_T  \sinh y_T \cosh y_L}
  =
  \frac{\epsilon \cosh y_T - \epsilon'}{k_T  \sinh y_T}
  \, .
\end{align}
The $\Phi$ integration with the delta-function yields
\begin{align}
  \left .
  \epsilon \frac{dN_m}{d^2 k_T dk_L}
  \right |_{k_L=0}
  &=
  \int_{-P_L^{\rm max}}^{P_L^{\rm max}} dP_L
  \int_{y_{T\rm min}}^{y_{T\rm max}}   dy_T 
  \,
  \sum_{i=\pm}
  E\frac{dN_M}{d^2P_T dP_L}(P_T, \Phi_i,P_L)
  \,
  \frac{1}{4\pi k_0}
  \frac{M}{k_T  |\sin \theta | }
  \label{eq:m-distribution}
\end{align}
with $\Phi_{\pm} -\phi = \pm \theta$.
Note that $y=y_{T\rm min}$ when $\P_T$ and $\k_T$ are parallel to each other
($\cos \theta=1$).

In a 2D model, we remove the $P_L$ integration and 
use the 2D decay distribution,
$\epsilon dn_m/d^2k_T = (1/2\pi) \delta (\epsilon_{\CM}-\epsilon_0)$,
instead.
The momentum distribution of the particle $m$ in the 2D model simplifies to
\begin{align}
  \epsilon \frac{dN_m}{d^2 k_T }
  &=
  \int_{y_{T\rm min}}^{y_{T\rm max}}  dy_T 
  \,
  \sum_{i=\pm}
  E \frac{dN_M}{d^2P_TdP_L}(P_T, \Phi_i, P_L=0)
\,
  \frac{1}{2\pi}
  \frac{M}{k_T | \sin \theta|}.
  \label{eq:m-distribution-2d}
\end{align}

\section{Thermal photons from the hydrodynamic model}
Here we give a brief explanation of the hydrodynamic model which we employ for calculating thermal photon production (see Figs.~\ref{fig:photon_centrality_rhic} and \ref{fig:pt_lhc}). 
We set the initial time of hydrodynamic expansion to $\tau_0 = 0.6$ fm/$c$, assuming that local thermal equilibrium is achieved by then.
A parametric initial condition model, 
TRENTo \cite{Moreland:2014oya,Ke:2016jrd},
is used to prepare the initial entropy density profile.
The parameter values are listed in Table~\ref{tab:param}.
We have re-tuned a normalization parameter $N$ and another parameter $p$ in TRENTo from comparison with experimental data, while keeping other parameters the same as those adopted in Ref.~\cite{Okamoto:2017rup}.  (The parameter $p$ interpolates between the IP-Glasma model $p=0$ and the wounded nucleon model $p=1$.)

We numerically solve the relativistic viscous hydrodynamic equation with a newly developed algorithm \cite{Akamatsu:2013wyk,Okamoto:2016pbc,Okamoto:2017ukz}.
Since we focus on heavy-ion collisions at RHIC’s highest and LHC energies, where the baryon number density in the created hot medium is close to zero, we utilize a parametrized equation of state based on lattice QCD calculations and hadron resonance  gas at zero baryon density \cite{Bluhm:2013yga}.
We allow for temperature dependence of shear and bulk viscosities, which we parametrize as~\cite{Niemi:2015qia,Denicol:2015nhu}
\begin{align}
  \frac{\eta}{s} (T)  & = \left ( \frac{\eta}{s}\right )_{\rm min} + c_1 (T_c -T) \theta(T_c -T) + c_2 (T-T_c) \theta(T-T_c)\, ,
  \label{eq:visco1}\\
  \frac{\zeta}{s} (T) &=b \frac{\eta}{s}(T) \left ( \frac{1}{3} - c_s^2 \right )^2 \, ,
\label{eq:visco2}
\end{align}
where $T_c$ is the pseudo-critical temperature set to $T_c = 167$ MeV \cite{Bluhm:2013yga} and $c_s$ is the sound velocity.
The value of $(\eta/s)_{\rm min}$ is 0.08, which is inferred from the AdS/CFT argument \cite{Kovtun:2004de}.
The parameters $c_1$, $c_2$, and $b$ are determined in Ref.~\cite{Okamoto:2017rup} so as to reproduce experimental data at the LHC.
Note that the value of $b$ 
is smaller than that of Ref.~\cite{Okamoto:2017rup} because we phenomenologically include here
the shear and bulk viscosity corrections on particle distributions in the ``particlization'' procedure
with the Cooper-Frye formula \cite{Bernhard:2018hnz}.
We switch the simulation scheme from a hydrodynamic model to a hadronic cascade model (we use UrQMD \cite{Bleicher:1999xi}) at temperature, $T_{\rm SW} = 150$ MeV, for studying the hadronic observables.
On the other hand, when we evaluate thermal photon production, we continue our hydrodynamic model simulation
in the hadronic phase down to the kinetic freeze-out temperature, $T_{\rm fo}=116$ MeV,
in order to obtain the temperature evolution profile, which is needed for the rate calculation.

\begin{table}
  \caption{Parameters of 3D relativistic viscous hydrodynamic model.
    \label{tab:param}}
  \begin{tabular}{ccccccccccccc}
    \hline
        & ~$N$~ & ~$p$~  & ~$k$~ & ~$w$~  & ~$\mu_0$~ & ~$\sigma_0$~ & ~$\gamma_0$~ & ~$J$~ & ~$(\eta/s)_{\rm min}$~ & ~$c_1$~ & ~$c_2$~ & ~$b$~ \\
    \hline
~RHIC~ & ~47~   & ~0.4~  & ~2.0~ & ~0.59~ & ~0.0~     & ~2.0~        & ~7.3~       & ~0.77~ & ~0.08~           & ~10~     & ~0.7~   & ~10~ \\
  LHC  & 103    & 0.2    & 2.0   & 0.59   & 0.0       & 2.7          & 7.3         & 0.74   & 0.08             & 10       & 0.7     & 10 \\
\hline
\end{tabular}
\end{table}

We assign the centrality class to the events based on initial entropy densities.
First we produce a certain number of minimum-bias events with initial conditions generated by TRENTo,
and then we arrange these events in decreasing order of magnitude of entropy
$\left . dS/dy \right |_{y=0} $. 
For the centrality class 0--10~\%, for example, 
we pick up the top 10~\% of the events with the large entropy in the entire sample set.
In the current calculation of thermal photon production with the relativistic viscous hydrodynamic model,
we prepared 4000 minimum-bias events each for RHIC and the LHC.

Thermal photons are emitted from QGP and hot hadronic matter phases during hydrodynamic expansion.
We evaluate two contributions by adopting the following smooth interpolation formula,
\begin{align}
  \epsilon \frac{dR_{\rm th}^\gamma}{d^3 k}
  = \frac{1}{2} \left ( 1- \tanh \frac{T-T_0}{\Delta T} \right )
  \epsilon \frac{dR_{\rm had}^\gamma}{d^3 k}
  +
  \frac{1}{2} \left ( 1 + \tanh \frac{T-T_0}{\Delta T} \right )
  \epsilon \frac{dR_{\rm QGP}^\gamma}{d^3 k} \, ,
\end{align}
where the interpolation parameters $T_0$ is set to $T_0 = 170$ MeV and $\Delta T = 0.1T_0$ \cite{Monnai:2019vup}.
We use the photon emission rate of QGP which parametrizes the result of the complete leading-order pQCD calculation
in the strong coupling constant $g_s$ \cite{Arnold:2001ms},
and the thermal photon emission rate in hadronic matter consisting of $\pi$, $\rho$ and $\omega$ mesons,
which is given in Refs.~\cite{Holt:2015cda,Turbide:2003si,Heffernan:2014mla}.

\section{
  Comparative study of photon $p_T$ spectrum and $v_2^\gamma$ with $\kappa$ parameter}

In the main text, we showed that the photon $p_T$ spectrum and elliptic flow coefficient $v_2^\gamma$ can be
reproduced simultaneously by including the photons from the radiative ReCo model.
By the fitting, we obtained the value of the $\kappa$ parameter,
which determines the size of the radiative ReCo contribution, to be 
$\kappa = 0.2$ and 0.05 for the RHIC and LHC data, respectively. 
These values are unnaturally large from the perspective of the small QED coupling constant $\alpha_{\rm em}=1/137$,
which may suggest some dynamical enhancement for radiative hadronization or
existence of other unknown photon source, among other possibilities.

Instead of fixing $\kappa$ parameter by the fitting to the data,
we vary $\kappa$ here to show how these observables depend on it.
In Fig.~\ref{fig:kappa-deps}, we plot (a) the photon $p_T$ spectrum  and
(b) elliptic flow $v_2^\gamma(p_T)$ for $\kappa =0.01$, 0.05, and 0.2 (from bottom to top in red solid lines),
together with the case without the radiative hadronization contribution $\kappa =0$,
{\it i.e.,} thermal+prompt photons (blue dashed line), for two centrality classes at the RHIC energy.
Note that the case $0.01 \sim O(\alpha_{\rm em})$ corresponds to the naive expectation.

In panel (a) we find that an additional photon contribution is needed
to fill the gap between the thermal+prompt photon spectrum and the observed data,
of size $\kappa = 0.2$ in the case of the radiative ReCo model.
The value $\kappa=0.05$ is insufficient to make up for this deficit,
and $\kappa=0.01$ yields only a negligible photon contribution compared to the thermal+prompt spectrum.
From panel (b) we also see that the flow coefficient $v_2^\gamma(p_T)$ of the thermal+prompt photons
is much smaller than the observed data, and we need a flow component of the photons
almost as strong as the pion flow. Again the radiative hadronization component with $\kappa=0.2$ gives rise
to a reasonable contribution to make $v_2^\gamma(p_T)$ consistent with the data at the RHIC energy,
while those with $\kappa = 0.01$ and 0.05 are insufficient.
At higher $p_T$ the prompt photon contribution dominates to suppress the $v_2^\gamma$ values.

It would be worthwhile to investigate a physical mechanism for the large $\kappa$ value in our model,
with keeping one's eyes open for any possibilities such as nonequilibrium dynamical effects,
viscous corrections to thermal contributions, and yet unknown photon sources.

\begin{figure}[bt] 
\begin{center}
\includegraphics[width=0.48\textwidth]{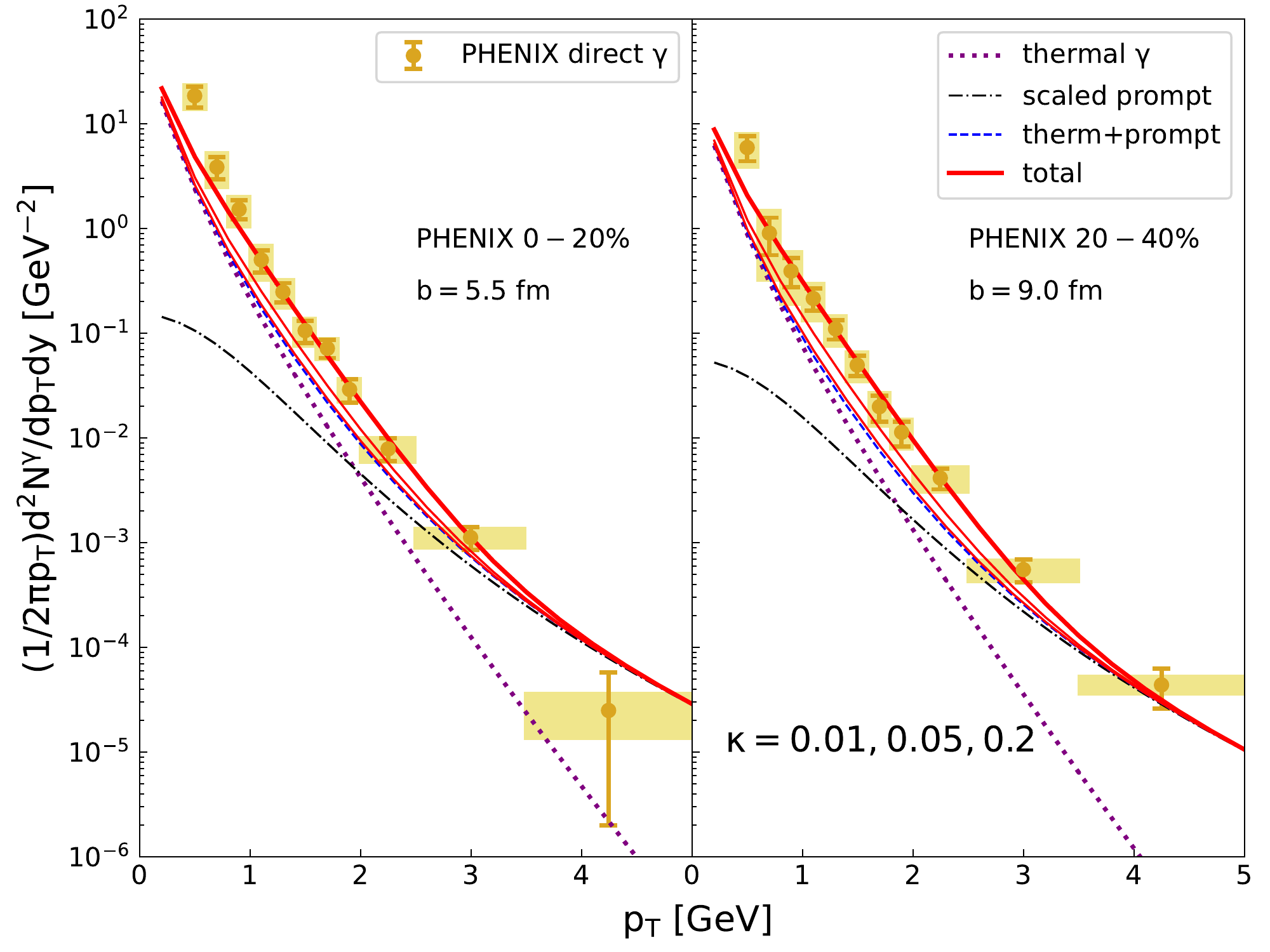}
\hfill
\includegraphics[width=0.48\textwidth]{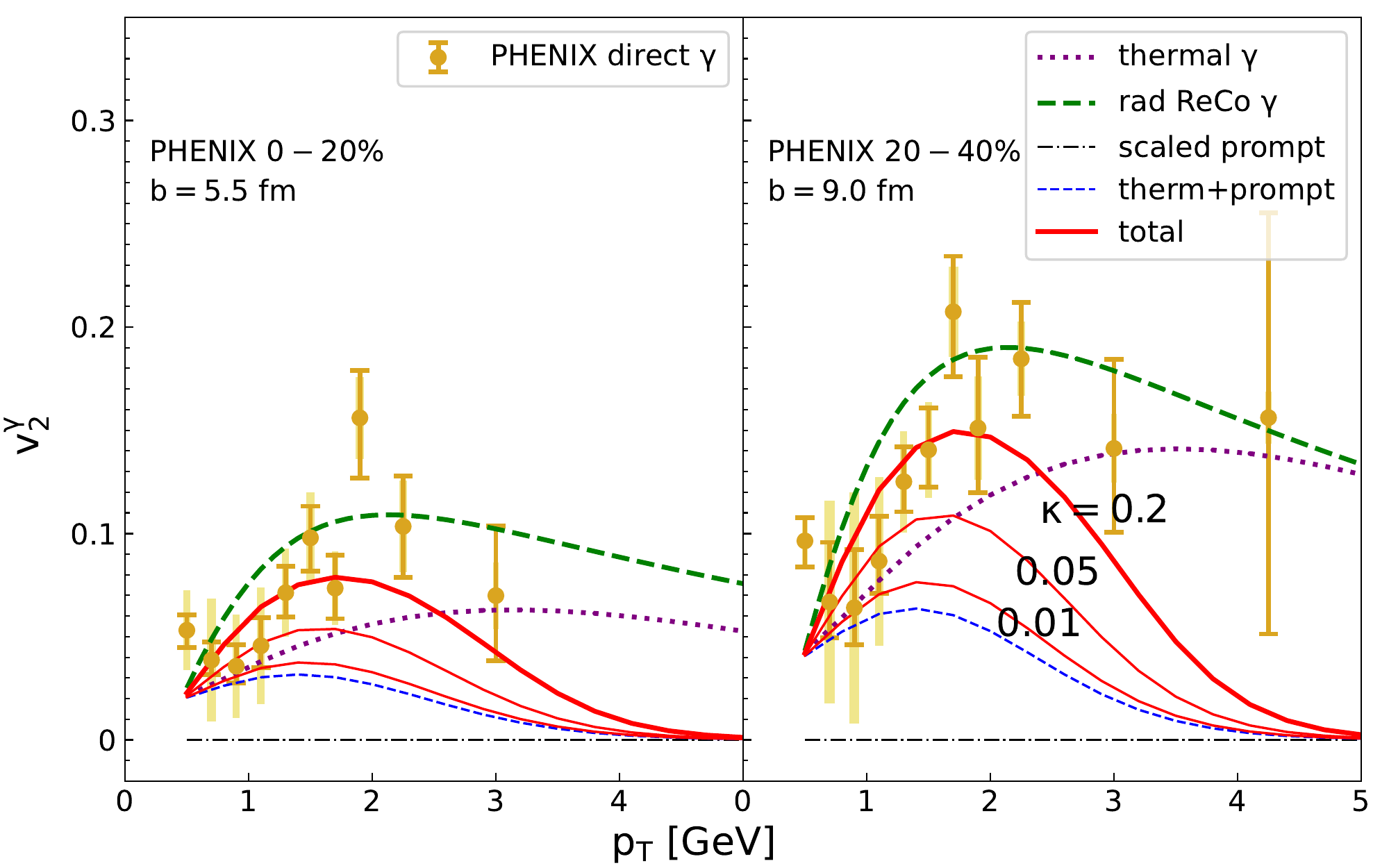}

\vspace{2mm}

{~}\hfill (a) \hfill\hfill (b) \hfill{~}
\\

\caption{Parameter $\kappa$ dependence of (a) photon $p_T$ spectrum and (b) elliptic flow coefficient $v_2^\gamma$.
  The solid curves represent the total photon spectrum and elliptic flow coefficient for $\kappa=0.01, 0.05$, and $0.2$, respectively,
  from bottom to top. Thermal photon (dotted) and prompt photon (dash-dotted) contributions and their sum (dashed) are also shown. In panel (b), the ellptic flow coefficent $v_2^\gamma$ of radiative hadronization photons is shown by the green long-dashed line.
\label{fig:kappa-deps}}
\end{center}
\end{figure}


\bibliography{photon}

\end{document}